\documentclass{aa}  
\usepackage{graphicx}
\usepackage{natbib}
\bibpunct{(}{)}{;}{a}{}{,}
\usepackage{txfonts}
\newcounter{Rco}
\newcommand{\Ionst}[1]{\setcounter{Rco}{#1}\Roman{Rco}}
\newcommand{\Ion}[2]{\mbox{#1\ {\scriptsize\Ionst{#2}}}}
\newcommand{\Ionw}[3]{\mbox{#1\ {\scriptsize\Ionst{#2}}~$\lambda\,#3$\,\AA}}

\newcommand{\Ionww}[3]{\mbox{#1\ {\scriptsize\Ionst{#2}}~$\lambda\lambda\,#3$\,\AA}}

\newcommand{\eq}[1]{\mbox{Equation\,\ref{#1}}}
\newcommand{\logg}{\mbox{$\log g$}}
\newcommand{\loggw}[1]{\mbox{$\log g\hspace{-0.5mm} =\hspace{-0.5mm}  #1$}}

\newcommand{\ab}[1]{\mbox{Fig.\,\ref{#1}}}
\newcommand{\sA}[1]{\mbox{(Fig.\,\ref{#1})}}
\newcommand{\ratio}[2]{\mbox{$n_{\rm #1}/n_{\rm #2}$}}

\newcommand{\se}[1]{\mbox{Sect.\,\ref{#1}}}
\newcommand{\sK}[1]{\mbox{(Sect.\,\ref{#1})}}
\newcommand{\sla}{\raisebox{-0.10em}{$\stackrel{<}{{\mbox{\tiny $\sim$}}}$}}

\newcommand{\ta}[1]{\mbox{Table\,\ref{#1}}}
\newcommand{\sT}[1]{\mbox{(Table\,\ref{#1})}}
\newcommand{\Teff}{\mbox{$T_\mathrm{eff}$}}
\newcommand{\Teffw}[1]{\mbox{$\Teff\hspace{-0.5mm}=\hspace{-0.5mm}#1\,\mathrm{kK}$}}
\newcommand{\lsv}{\object{LS\,V\,$+46\degr 21$}}
\newcommand{\pnsh}{\object{Sh\,2$-$216}}
\renewcommand{\object}{{}}
\begin{document}
\title{High-resolution FUSE and HST ultraviolet spectroscopy\\
       of the white dwarf central star of \pnsh
        \thanks
        {Based on observations with the NASA/ESA Hubble Space Telescope, obtained at the Space Telescope Science 
         Institute, which is operated by the Association of Universities for Research in Astronomy, Inc., under 
         NASA contract NAS5-26666.
        }
        \thanks
        {Based on observations made with the NASA-CNES-CSA Far Ultraviolet 
         Spectroscopic Explorer. FUSE is operated for NASA by the Johns Hopkins
         University under NASA contract NAS5-32985.
        }
\author{T\@. Rauch\inst{1}         \and 
        M\@. Ziegler\inst{1}       \and 
        K\@. Werner\inst{1}        \and 
        J\@. W\@. Kruk\inst{2}     \and 
        C\@. M\@. Oliveira\inst{2} \and
        D\@. Vande Putte\inst{3}   \and
        R\@. P\@. Mignani\inst{3}  \and
        F\@. Kerber\inst{4}
       }
}

\institute{Institut f\"ur Astronomie und Astrophysik, Universit\"at T\"ubingen, Sand 1, 72076 T\"ubingen, Germany
           \and
           Department of Physics and Astronomy, Johns Hopkins University, Baltimore, MD 21218, U.S.A.
           \and
           Mullard Space Science Laboratory, University College London, Holmbury St Mary, Dorking, Surrey RH5 6NT, United Kingdom
           \and
           European Southern Observatory, Karl-Schwarzschild-Stra\ss e 2, 85748 Garching, Germany
           }

\offprints{T\@. Rauch,\\ \email{rauch@astro.uni-tuebingen.de}}

\date{Received 25 January, 2007; accepted 11 April 2007}


\abstract
         {We perform a comprehensive spectral analysis of \lsv\ 
          in order to compare its photospheric properties to theoretical predictions from
          stellar evolution theory as well as from diffusion calculations.
         }
         {\lsv\ is the DAO-type central star of the planetary nebula \pnsh.
          High-resolution, high-S/N ultraviolet observations obtained with FUSE and STIS aboard the HST
          as well as the optical spectrum  
          have been analyzed in order to determine the photospheric parameters
          and the spectroscopic distance.}
         {We performed a detailed spectral analysis of the ultraviolet and optical spectrum
          by means of state-of-the-art NLTE model-atmosphere techniques.}
         {From the 
          \Ion{N}{4} -- \Ion{N}{5}, 
          \Ion{O}{4} -- \Ion{O}{6}, 
          \Ion{Si}{4} -- \Ion{Si}{5}, and 
          \Ion{Fe}{5} -- \Ion{Fe}{7} 
          ionization equilibria, 
          we determined an effective temperature of $(95\pm 2)\,\mathrm{kK}$ with high precision.
          The surface gravity is \loggw{6.9\pm 0.2}.
          An unexplained discrepancy appears between the   
          spectroscopic distance $d = 224^{+46}_{-58}\,\mathrm{pc}$ and
          the parallax distance $d = 129^{+6}_{-5 }\,\mathrm{pc}$ of \lsv.
          For the first time, 
          we have identified \Ion{Mg}{4} and \Ion{Ar}{6} absorption lines 
          in the spectrum of a hydrogen-rich central star and 
          determined the Mg and Ar abundances as well as
          the individual abundances of iron-group elements (Cr, Mn, Fe, Co, and Ni).
          With the realistic treatment of metal opacities up to the iron group in the model-atmosphere calculations,
          the so-called Balmer-line problem (found in models that neglect metal-line blanketing) vanishes.}
         {Spectral analysis by means of NLTE model atmospheres has presently arrived at a high level of sophistication,
          which is now hampered largely by the lack of reliable atomic data and accurate line-broadening tables. 
          Strong efforts should be made to improve upon this situation.
         } 


\maketitle

\section{Introduction}
\label{sect:introduction}

The planetary nebula (PN) \pnsh\ (\object{PN\,G158.6+00.7})
has been discovered as a large and faint emission nebula (\object{Sharpless 216})
and was classified as an \Ion{H}{2} region.
\citet{f1981} performed spectrophotometry on this ``curious emission-line nebula''
and found several forbidden lines and properties similar to PNe.
\citet{r1985} used high-resolution Fabry-Perot spectrometry to show that 
\pnsh\ has all the characteristics of an extremely old PN with 
a very low expansion velocity $v_\mathrm{exp} < 4\,\mathrm{km\,s^{-1}}$, 
but a central star (CS) was not found. 
\citet{r1985} speculated that the CS is no longer centrally located due to
deceleration of the PN shell over a long period of time by the interstellar medium (ISM).

\citet{cr1985} have unambiguously identified \lsv, one of two possible CS candidates
(the other was \object{AS\,84} \citep*{htv1965}) located nearly midway between the
apparent center of \pnsh\ and its eastern rim, 
as the exciting star of \pnsh\ by proper-motion measurements.  
\pnsh\ has obviously experienced a mild interaction with the ISM \citep*{tmn1995}. 
From its distance and proper motion, \citet*{kea2004} have determined that it has a 
thin-disk orbit of low inclination and eccentricity, and that the CS left the center of its 
surrounding PN about 45\,000 years ago.

At a distance of $d=129\,\mathrm{pc}$ \citep{hea2007}, it is the closest possible PN known. 
With an apparent size of $100'\,\times\,100'$ \citep{bea1990}, it is among the physically most extended
\citep[cf\@.][]{rea2004} and hence, oldest PNe \citep[about 660\,000 years,][]{n1999} known.

Since its identification as the CS of \pnsh, \lsv\ has been object of many investigations and 
analyses which are briefly summarized here.

\citet{fb1990} were able to detect a large number of \Ion{Fe}{5} and \Ion{Fe}{6} lines in spectra taken by the
International Ultraviolet Explorer (IUE).
\citet{tn1992} have demonstrated that \lsv, which is the brightest \citep[$m_\mathrm{V} = 12.67\pm 0.02$,][]{cea1993} 
DAO-type white dwarf (\object{WD\,0441+467}, \citet*{ms1999}) known, 
has the properties (\Teffw{90}, \loggw{7} ($\mathrm{cm\,s^{-2}}$), $\mathrm{He/H} = 0.01$ by number)
necessary to ionize the surrounding nebula. 

\citet{n1992, n1993}, \citet{ns1993}, and \citet{nr1994} have reported that the Balmer-series lines of
very hot DAO white dwarfs could not all be fit simultaneously with an NLTE model of a given \Teff.  
For example, fits to the individual Balmer lines of \lsv\ gave values of \Teff\ ranging from about 50\,kK
for H$\,\alpha$ to 90\,kK for H$\,\delta$.
This circumstance is known as the Balmer line problem (BLP).
\citet{brl1993} found (in LTE calculations for DAO WDs) that the BLP is reduced by the consideration of 
metal-line blanketing, and \citet{bea1994} could show clearly that the presence of heavy metals is the source of the BLP.
\citet*{bhh1998} have shown that models which neglect the opacity of heavy elements are not well suited 
for the analysis of DA white dwarfs.
To summarize, pure hydrogen models are not suited for the spectral analysis of hot DA(O) WDs in general
(cf\@. \ab{fig:tefffe}).

\citet{w1996} calculated NLTE model atmospheres for \lsv\ based on parameters of \citet{tn1992}
and introduced C, N, and O (at solar abundances) in addition. Surface cooling by these metals as well as the detailed
consideration of the Stark line broadening in the model-atmosphere calculation has the effect that the
BLP almost vanishes in \lsv. Later, \citet{kw1998} could demonstrate that these model atmospheres
reproduce well HUT (Hopkins Ultraviolet Telescope) observations of \lsv\ within $912 - 1840\,\mathrm{\AA}$
at $T_\mathrm{eff}=85\,\mathrm{kK}$ and $\log g = 6.9$.

\citet{n1999} determined \Teffw{83.2\pm 3.3} and \loggw{6.74\pm 0.19}. 
The model contained only H and He (at an abundance ratio of $\log$\,\ratio{He}{H}$=-1.95$)
and a fit to H$\delta$ was done to derive \Teff\ (cf\@. \ab{fig:tefffe}).
Also, instead of assuming LTE, which fails in the region of typical white dwarf central stars (\Teffw{100} / \loggw{7.0}),
he calculated an NLTE model based on the accelerated lambda iteration (ALI) method as described by \citet{w1986}. 

\citet{wea2003a} evaluated 
the \Ion{Fe}{5} -- \Ion{Fe}{6} ionization equilibrium and arrived at \Teffw{90} and \loggw{7.0}. 

\citet{tea2005} considered the opacity of light metals (C, N, O, and Si) in addition and determined
\Teffw{93\pm 5} and \loggw{6.9\pm 0.2} ($\mathrm{cm\,s^{-1}}$).
\Teff\ and \logg\ were derived from the ionization equilibrium of \Ion{O}{4} -- \Ion{O}{6}. 
The abundances for the included elements were determined to be
[He]\,=\,\mbox{$-$0.9},
 [C]\,=\,\mbox{$-$1.0}, 
 [N]\,=\,\mbox{$-$2.0},
 [O]\,=\,\mbox{$-$0.9}, and
[Si]\,=\,\mbox{$-$0.3}
([x]: log (mass fraction / solar mass fraction) of species x).
From \Ionw{O}{5}{1371} a radial velocity of $v_\mathrm{rad} = +22.4\,\mathrm{km\,s^{-1}}$ was measured.   

Recently, \citet{h2005} determined an oversolar iron abundance in a preliminary analysis based on
H+He+Fe models.

In this paper, we present a detailed analysis of the individual abundances of iron-group elements 
and light metals based on
high-resolution UV observations \sK{sect:obs}. 
The analysis is described in \se{sect:modeling}.

\section{Observations and reddening}
\label{sect:obs}

Spectral analysis by model-atmosphere techniques needs observations
of lines of successive ionization stages in order to evaluate the ionization equilibrium of a particular
species which is a sensitive indicator of $T_\mathrm{eff}$. 
For stars with \Teff\ as high as $\approx$90\,kK, the ionization degree is very high and most 
of the metal lines are found at UV wavelengths.
Thus, high-S/N and high-resolution UV spectra
are a prerequisite for a precise analysis. Consequently, we used 
FUSE (Far Ultraviolet Spectroscopic Explorer) 
and
HST/STIS (Space Telescope Imaging Spectrograph aboard the Hubble Space Telescope)
in order to obtain suitable data.

\subsection{The FUSE spectrum of \lsv}
\label{subsect:fuse}

FUSE provides spectra in the wavelength band 900\,--\,1187\,\AA\ with a typical resolving power
of 20\,000. FUSE consists of four independent co-aligned telescopes with prime-focus
Rowland circle spectrographs. Two of the four channels have optics coated
with Al$+$LiF and two channels have optics coated with SiC.
Each spectrograph has three entrance apertures:  LWRS (30\arcsec $\times$ 30\arcsec),
MDRS (4\arcsec $\times$ 20\arcsec), and HIRS (1\farcs 25 $\times$ 20\arcsec).
Further information on the FUSE mission and instrument can be found in \citet{mea2000} and \citet{sea2000}.

This star was observed many times in all three spectrograph apertures (LWRS, MDRS, and HIRS)
as part of the wavelength calibration program for FUSE. The observations used in this analysis
are listed in \ta{tab:logfuse}. The LWRS observations were photometric or nearly so in all four
channels, so the effective exposure time was essentially the same for each of them.
The effective exposure time varied considerably in the MDRS channels as a result of
channel misalignments. The time given in \ta{tab:logfuse} for these observations is that of the
LiF1 channel; the LiF2 time is similar but the SiC channel exposure time is typically lower by a
factor of two.

\begin{table}[ht]
\caption[]{FUSE observations of \lsv\ used in this analysis.} 
\label{tab:logfuse}
\begin{tabular}{cccc}
\hline\hline
\noalign{\smallskip}
dataset & start time (UT) & aperture & exp time (sec) \\
\hline
\noalign{\smallskip}
M1070404 & 2001-01-10 15:29 & LWRS & 5606 \\
M1070407 & 2001-01-25 14:46 & LWRS & 6826 \\
M1070416 & 2003-02-06 09:59 & LWRS & 7511 \\
M1070419 & 2003-09-27 00:48 & LWRS & 3541 \\
M1070422 & 2004-01-23 17:40 & LWRS & 2990 \\
\hline
\noalign{\smallskip}
M1070402 & 2001-01-09 17:36 & MDRS & 1418 \\
M1070405 & 2001-01-23 16:15 & MDRS & 8922 \\
M1070408 & 2001-01-25 19:58 & MDRS & 6257 \\
M1070417 & 2003-02-06 15:05 & MDRS & 4899 \\
M1070420 & 2003-09-27 15:55 & MDRS & 3613 \\
M1070423 & 2004-01-26 06:49 & MDRS & 4401 \\
\hline
\end{tabular}
\end{table}

The raw data for each exposure were reduced using CalFUSE version 3.0.7.
For a description of the CalFUSE pipeline, see \citet{dea2007}.
The exposure with the maximum flux outside of airglow lines was then identified for
each observation. Any exposures with less than 40\% of this flux were discarded and
the remainder were normalized to match the peak exposure and their exposure times were
scaled accordingly. These corrections were negligible (less than 1\%) for all but
the last exposure of M1070422, where a delayed target acquisition resulted in a significant
correction. For the MDRS spectra these corrections were negligible for the LiF channels
but were typically $\sim$50\% for the SiC channels. The MDRS spectra were subsequently
renormalized to match the LWRS spectra. The exposures were coaligned
by cross-correlating on narrow ISM features and combined. This coalignment was then
repeated to combine the observations; the end result was a single spectrum for each channel
and spectrograph aperture.
Subsequent analyses were performed primarily using the LWRS spectra, with the MDRS spectra
providing a confirmation that weak features were real and not the result of detector fixed-pattern
noise.

At first glance, the FUSE spectrum of \lsv\ gives the impression that it is hopeless to find signatures of the
stellar photosphere in the sea of interstellar absorption lines \sA{fig:nhi}. 
However, some strong, isolated lines, e.g\@. of \Ion{N}{4}, \Ion{P}{5}, \Ion{S}{6}, and \Ion{Cr}{6}, as well as
a large number of \Ion{Fe}{6}, \Ion{Fe}{7}, and \Ion{Ni}{6} lines, are identified and are included in our analysis.

In order to demonstrate that a combined model of photospheric and interstellar absorption
reproduces the FUSE observation well, we decided to apply the profile-fitting procedure 
{\sc OWENS} which enables us to model the interstellar absorption lines with high accuracy. 
The use of this procedure is described in \se{sect:owens}.

\subsection{The STIS spectrum of \lsv}
\label{subsect:stis}

The STIS observation had been performed with the E140M grating and the FUV-MAMA detector which provides 
\'echelle spectra in the wavelength range from 1144\,\AA\ to 1729\,\AA\ with a theoretical
resolving power of $\lambda/\Delta\lambda = 45\,800$). 
Two spectra (total exposure time 5.5\,ksec, resolution $\approx 0.06$\,\AA) were taken in
2000 and processed by the standard pipeline data reduction (version of June 2006).

In order to increase the signal-to-noise ratio ($S/N$), the two obtained spectra have been co-added and 
were subsequently smoothed with a Savitzky-Golay filter \citep{sg1964}.
This spectrum has a very good $S/N > 50$.


\subsection{Interstellar absorption and reddening}
\label{subsect:ISMbasic}

In order to fine tune model-atmosphere parameters and better match the observations, 
we have determined from all models both the column
density of interstellar neutral hydrogen ($n_\ion{H}{i}$) and the reddening ($E_\mathrm{B-V}$).

\begin{figure}[ht]
  \resizebox{\hsize}{!}{\includegraphics{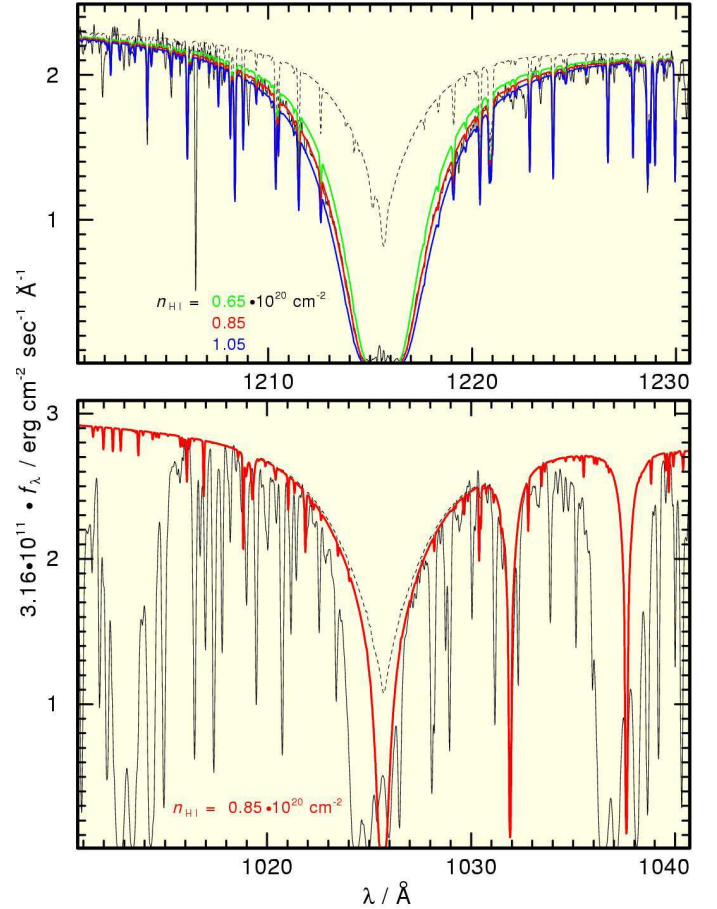}}
  \caption[]{Synthetic spectra around
             \Ion{H}{1} Ly\,$\alpha$ (top) and 
                        Ly\,$\beta$ (bottom)
             calculated with different $n_\ion{H}{i}$
             compared with sections of the STIS and FUSE spectra of \lsv, respectively.
             Note the strong interstellar absorption features in the lower panel.
             An additional factor of 0.85 is applied in order to normalize the synthetic stellar spectrum
             to the observation at 1020\,\AA\ and 1030\,\AA. 
             The blue wing of Ly\,$\beta$ is blended with interstellar H$_2$ ($J=0,1$) absorption.
             The observation of Ly\,$\alpha$ is best matched at 
             $n_\ion{H}{i} = 8.5\times 10^{19}\,\mathrm{cm^{-2}}$.
             The thin, dashed spectra are calculated without interstellar \Ion{H}{1} absorption.
             For the model-atmosphere parameters, see \se{sect:conclusions}.
             All synthetic spectra shown in this paper are normalized to match the observation
             at 1700\,\AA\ and are convolved with Gaussians in order to match the
             instrument resolution \sK{sect:obs}. 
             The observed spectrum is normalized by a factor of $3.16\times 10^{11}$.
            }
  \label{fig:nhi}
\end{figure}

From a detailed comparison of 
\Ion{H}{1} L\,$\alpha$  \sA{fig:nhi}
with the observation (best suited because the interstellar \Ion{H}{1} absorption dominates
the complete line profile), 
we determined $N_\mathrm{H\,I} = (8.5 \pm 1.0) \times 10^{19}\,\mathrm{cm^{-2}}$.  
For the following analysis, we adopt this value.

We determine the reddening from the FUSE and STIS spectra \sA{fig:ebv}.
We note that the FUSE and STIS fluxes agree quite well for $1150 - 1180$\,\AA.
We achieve the best match to the continuum slope with $E_\mathrm{B-V} = 0.065^{+0.010}_{-0.015}$ \sA{fig:ebv}

\begin{figure*}[ht]
  \resizebox{\hsize}{!}{\includegraphics{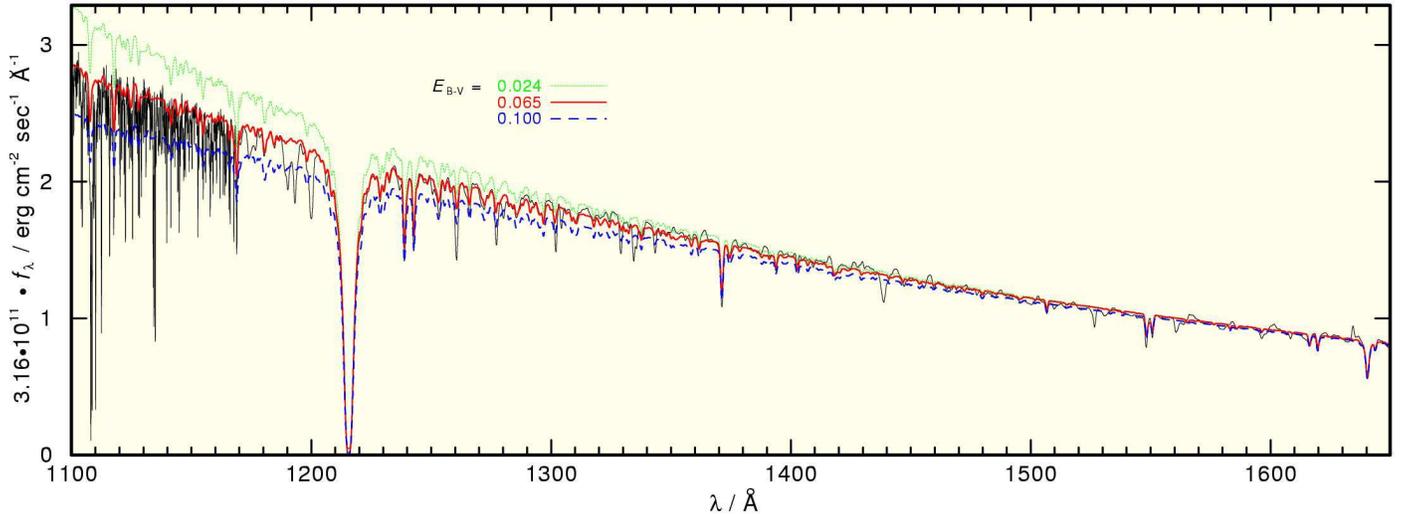}}
  \caption[]{Sections of the FUSE ($1150 - 1170$\,\AA) and STIS ($1170 - 1650$\,\AA) spectra of \lsv\ compared with 
             synthetic spectra at different $E_\mathrm{B-V}$ (for clarity, the FUSE and STIS spectra are smoothed with 
             Gaussians of 0.1\,\AA\ and 1\,\AA\ FWHM, respectively). The synthetic spectra are 
             smoothed with Gaussians of 1\,\AA\ FWHM and are normalized to
             the flux level of the observation around \Ionw{He}{2}{1640}.
             Due to the many (photospheric and interstellar) absorption lines,
             a convolution of the FUSE spectrum with a wider Gaussian would result in
             an artificially lower ``continuum''.
             The overall continuum is well reproduced at $E_\mathrm{B-V} = 0.065$.
            }
  \label{fig:ebv}
\end{figure*}

This value is higher than the $E_\mathrm{B-V} = 0.024\pm 0.008$ predicted by the Galactic reddening law of \citet{gl1989}. 
In the case of \lsv\ its PN is likely to modify the expect reddening behaviour in two ways: 
first, the enormous ambient PN will result in additional reddening (although the column density is small), and 
second, presence of the circumstellar matter is probably modifying the applied reddening law \citep{s1979}. 
Our $E_\mathrm{B-V}$ is lower than  $E_\mathrm{B-V} = 0.1$ that \citet{kw1998} 
derived from the analysis of a HUT spectrum ($915\,\mathrm{\AA} < \lambda < 1840\,\mathrm{\AA}$).
Recently, \citet{hea2007} have determined $E_\mathrm{B-V} = 0.08$ from optical spectra.

We have to mention that fitting the extinction is uncertain. 
The generic curves are averages to many sight lines and often don't fit a single one very well.
The curves are less well-characterized as one goes farther into the FUV, 
as there are multiple absorbers possible and they each have a distinctive wavelength-dependence
\citep[see, e.g.,][]{sea2005}. 
However, the exact $E_\mathrm{B-V}$ is not important, neither for our following spectral analysis by
detailed line-profile fitting \sK{subsect:analysis} 
nor for our distance determination \sK{subsect:MdL}. 
We finally adopt $E_\mathrm{B-V} = 0.065$.

\section{Data modeling and analysis}
\label{sect:modeling}

In the following, we describe in detail the analysis of the FUSE and STIS spectra of \lsv\ by means of
NLTE model-atmosphere techniques.

\subsection{Model atmospheres and atomic data}
\label{subsect:model}

We employed {\it TMAP}, the T\"ubingen NLTE Model Atmosphere Package
\citep{wea2003b}, for the calculation of plane-parallel, homogeneous, static models which consist
of H, He, C, N, O, F, Mg, Si, P, S, Ar, Ca, Sc, Ti, V, Cr, Mn, Fe, Co, and Ni (in the
following, we will refer to Ca -- Ni as iron-group elements). H, He, C, N, O, F, Mg, Si, P, S, and Ar
are represented by ``classical'' model atoms \citep{r1997} partly taken from {\it TMAD}, 
the T\"ubingen Model Atom Database\footnote{http://astro.uni-tuebingen.de/\raisebox{.2em}{\small $\sim$}rauch/TMAD/TMAD.html}.

For Ca+Sc+Ti+V+Cr+Mn+Fe+Co+Ni individual model atoms are constructed by
{\it IrOnIc} \citep*{rd2003}, using a statistical approach in order to treat the
extremely large number of atomic levels and line transitions by the introduction
of ``super-levels'' and ``super-lines''. In total 686 levels are treated in NLTE,
combined with 2417 individual lines and about 9 million iron-group lines \sT{tab:statistics}, 
taken from \citet{k1996} as well as from the OPACITY and IRON projects \citep{sea1994,hea1993}.
It is worthwhile to mention that this is the most extended model we have calculated with
{\it TMAD} so far. The frequency grid comprises about 34\,000 points spanning 
$10^{12} - 3\times 10^{17}$\,Hz.
On a PC of our institute's cluster with a 3.4\,GHz processor and 4\,GB RAM,
one iteration takes about 48\,000\,s. 
We need more than one week of CPU time to calculate one complete model
with a convergence criterion of $10^{-4}$ in the relative corrections of temperature, densities,
occupation numbers, and flux for all (90) depth points. 

\begin{table}[ht]
\caption{Statistics of ``classical`` (left) and iron-group (Ca -- Ni, right) model atoms used in our calculations.
         We give the number of levels treated in NLTE and the respective line transitions.
         For the iron-group levels, we list the number of so-called ``super levels''
         and sampled lines that are statistically combined to ``super lines''.}         
\label{tab:statistics}
\begin{tabular}{lrrllrrr}
\hline
\hline
\noalign{\smallskip}
             &        &       & &              & super  & super &  sample \\
ion          & levels & lines & & ion          & levels & lines &  lines  \\
\cline{1-3}
\cline{5-8}
\noalign{\smallskip}
\Ion{H}{1}   &     10 &    45 & & \Ion{Ca}{5}  &      3 &    6  &  141956 \\ 
\Ion{H}{2}   &      1 &     - & & \Ion{Ca}{6}  &      7 &   24  &  114545 \\
\Ion{He}{1}  &      5 &     3 & & \Ion{Ca}{7}  &      7 &   27  &   71608 \\ 
\Ion{He}{2}  &     14 &    90 & & \Ion{Ca}{8}  &      1 &    0  &       0 \\
\Ion{He}{3}  &      1 &     - & & \Ion{Sc}{5}  &      3 &    6  &   65994 \\
\Ion{C}{3}   &     10 &    12 & & \Ion{Sc}{6}  &      7 &   24  &  237271 \\
\Ion{C}{4}   &     54 &   295 & & \Ion{Sc}{7}  &      7 &   26  &  176143 \\
\Ion{C}{5}   &      1 &     0 & & \Ion{Sc}{8}  &      1 &    0  &       0 \\
\Ion{N}{4}   &      9 &    10 & & \Ion{Ti}{5}  &      3 &    5  &   26654 \\
\Ion{N}{5}   &     54 &   297 & & \Ion{Ti}{6}  &      7 &   24  &   95448 \\
\Ion{N}{6}   &      1 &     0 & & \Ion{Ti}{7}  &      7 &   26  &  230618 \\
\Ion{O}{4}   &     11 &    19 & & \Ion{Ti}{8}  &      1 &    0  &       0 \\
\Ion{O}{5}   &     90 &   608 & & \Ion{V}{5}   &      3 &    5  &    2123 \\
\Ion{O}{6}   &     54 &   291 & & \Ion{V}{6}   &      7 &   24  &   35251 \\
\Ion{O}{7}   &      1 &     0 & & \Ion{V}{7}   &      7 &   24  &  112883 \\
\Ion{F}{5}   &      8 &     9 & & \Ion{V}{8}   &      1 &    0  &       0 \\
\Ion{F}{6}   &      6 &     4 & & \Ion{Cr}{5}  &      3 &    5  &   43860 \\
\Ion{F}{7}   &      2 &     1 & & \Ion{Cr}{6}  &      7 &   24  &    4406 \\
\Ion{F}{8}   &      1 &     0 & & \Ion{Cr}{7}  &      7 &   24  &   37070 \\
\Ion{Mg}{3}  &      1 &     0 & & \Ion{Cr}{8}  &      1 &    0  &       0 \\
\Ion{Mg}{4}  &      8 &     7 & & \Ion{Mn}{5}  &      3 &    5  &  285376 \\
\Ion{Mg}{5}  &      5 &     2 & & \Ion{Mn}{6}  &      7 &   24  &   70116 \\
\Ion{Mg}{6}  &      1 &     0 & & \Ion{Mn}{7}  &      7 &   24  &    8277 \\
\Ion{Si}{3}  &      6 &     4 & & \Ion{Mn}{8}  &      1 &    0  &       0 \\
\Ion{Si}{4}  &     16 &    44 & & \Ion{Fe}{5}  &      3 &    5  &  793718 \\
\Ion{Si}{5}  &     15 &    20 & & \Ion{Fe}{6}  &      7 &   25  &  340132 \\
\Ion{Si}{6}  &      1 &     0 & & \Ion{Fe}{7}  &      7 &   24  &   86504 \\
\Ion{P}{3}   &      3 &     0 & & \Ion{Fe}{8}  &      1 &    0  &       0 \\
\Ion{P}{4}   &     15 &     9 & & \Ion{Co}{5}  &      3 &    5  & 1469717 \\
\Ion{P}{5}   &     18 &    12 & & \Ion{Co}{6}  &      7 &   22  &  898484 \\
\Ion{P}{6}   &      1 &     0 & & \Ion{Co}{7}  &      7 &   23  &  492913 \\
\Ion{S}{4}   &      6 &     4 & & \Ion{Co}{8}  &      1 &    0  &       0 \\
\Ion{S}{5}   &     14 &    16 & & \Ion{Ni}{5}  &      3 &    5  & 1006189 \\
\Ion{S}{6}   &     18 &    48 & & \Ion{Ni}{6}  &      7 &   22  & 1110584 \\
\Ion{S}{7}   &      1 &     0 & & \Ion{Ni}{7}  &      7 &   22  &  688355 \\
\Ion{Ar}{4}  &      2 &     0 & & \Ion{Ni}{8}  &      1 &    0  &       0 \\
\Ion{Ar}{5}  &      9 &     6 & &              &        &       &         \\
\Ion{Ar}{6}  &     15 &    21 & &              &        &       &         \\
\Ion{Ar}{7}  &     20 &    36 & &              &        &       &         \\
\Ion{Ar}{8}  &     13 &    24 & &              &        &       &         \\
\Ion{Ar}{9}  &      1 &     0 & &              &        &       &         \\
\cline{2-3}
\cline{6-8}
\noalign{\smallskip}                    
total        &    522 &   1937 & &              &    162 &   480 & 8646195 \\
\hline
\end{tabular}
\end{table}

\subsection{Line identification}
\label{subsect:lines}

The UV spectrum of \lsv\ exhibits a large number of stellar and interstellar absorption features.
None of the identified photospheric lines shows evidence for on-going mass loss of \lsv.
 
We are able to identify and reproduce about 95\% of all spectral
lines in the FUSE and STIS spectra of \lsv. It is likely that most of the remaining unidentified lines 
(e.g\@. in \ab{fig:tefffe})
stem from the same ions, but are from transitions whose wavelengths are not sufficiently well-known.
For instance for \ion{Fe}{vii}, \citet{k1996} provides only 22 laboratory measured (POS) lines and 1952 lines
with theoretical line positions (LIN lines). This situation is even worse for other ions (\ion{Fe}{vi}: 224 and 58664,
respectively) and species.

Despite these problems, we identify weak \Ion{Mg}{4} lines in the STIS spectrum \sT{tab:mgsiar}.
The strongest of which, \Ionw{Mg}{4}{1683.0}, is shown in \ab{fig:mgsiident}.
\citet{nd1975} had already proposed this identification for a respective feature in
the supergiant \object{$\delta$\,Ori}.

\begin{table}[ht]
\caption{Identified Mg, Si, and Ar lines in the STIS spectrum of \lsv.}         
\label{tab:mgsiar}
\begin{tabular}{lr@{.}lllcll}
\hline
\hline
\noalign{\smallskip}
ion & \multicolumn{2}{c}{$\lambda / \mathrm{\AA}$} &  \multicolumn{5}{c}{transition} \\
\hline         
\noalign{\smallskip}
\Ion{Mg}{4} & 1336&850 & 3p       & $\mathrm{^4 P^{o}_{5/2}}$   &--& 3d       & $\mathrm{^4 D^{~}_{3/2}}$   \\
\Ion{Mg}{4} & 1342&163 & 3p       & $\mathrm{^4 P^{o}_{5/2}}$   &--& 3d       & $\mathrm{^4 D^{~}_{5/2}}$   \\
\Ion{Mg}{4} & 1346&543 & 3p       & $\mathrm{^4 P^{o}_{5/2}}$   &--& 3d       & $\mathrm{^4 D^{~}_{7/2}}$   \\
\Ion{Mg}{4} & 1352&020 & 3p       & $\mathrm{^4 P^{o}_{3/2}}$   &--& 3d       & $\mathrm{^4 D^{~}_{5/2}}$   \\
\Ion{Mg}{4} & 1387&494 & 3p       & $\mathrm{^4 D^{o}_{5/2}}$   &--& 3d       & $\mathrm{^4 F^{~}_{7/2}}$   \\
\Ion{Mg}{4} & 1437&610 & 3p       & $\mathrm{^2 D^{o}_{5/2}}$   &--& 3d       & $\mathrm{^2 F^{~}_{7/2}}$   \\
\Ion{Mg}{4} & 1490&433 & 3s       & $\mathrm{^4 P^{~}_{3/2}}$   &--& 3p       & $\mathrm{^4 S^{o}_{3/2}}$   \\
\Ion{Mg}{4} & 1520&967 & 3p       & $\mathrm{^4 S^{o}_{3/2}}$   &--& 3d       & $\mathrm{^4 P^{~}_{1/2}}$   \\
\Ion{Mg}{4} & 1658&851 & 3s       & $\mathrm{^4 P^{~}_{5/2}}$   &--& 3p       & $\mathrm{^4 D^{o}_{5/2}}$   \\ 
\Ion{Mg}{4} & 1669&574 & 3s       & $\mathrm{^4 P^{~}_{3/2}}$   &--& 3p       & $\mathrm{^4 D^{o}_{1/2}}$   \\
\Ion{Mg}{4} & 1679&960 & 3s       & $\mathrm{^4 P^{~}_{3/2}}$   &--& 3p       & $\mathrm{^4 D^{o}_{3/2}}$   \\
\Ion{Mg}{4} & 1683&003 & 3s       & $\mathrm{^4 P^{~}_{5/2}}$   &--& 3p       & $\mathrm{^4 D^{o}_{7/2}}$   \\
\Ion{Mg}{4} & 1692&675 & 3s       & $\mathrm{^4 P^{~}_{1/2}}$   &--& 3p       & $\mathrm{^4 D^{o}_{1/2}}$   \\
\Ion{Mg}{4} & 1698&784 & 3s       & $\mathrm{^4 P^{~}_{3/2}}$   &--& 3p       & $\mathrm{^4 D^{o}_{5/2}}$   \\
\Ion{Mg}{4} & 1703&357 & 3s       & $\mathrm{^4 P^{~}_{1/2}}$   &--& 3p       & $\mathrm{^4 D^{o}_{3/2}}$   \\ 
\noalign{\smallskip}
\Ion{Si}{4} & 1210&652 & 4s       & $\mathrm{^{2} S^{~}_{1/2}}$ &--& 5p       & $\mathrm{^{2} P^{o}_{3/2}}$ \\
\Ion{Si}{4} & 1211&757 & 4s       & $\mathrm{^{2} S^{~}_{1/2}}$ &--& 5p       & $\mathrm{^{2} P^{o}_{1/2}}$ \\
\Ion{Si}{4} & 1402&770 & 3s       & $\mathrm{^{2} S^{~}_{1/2}}$ &--& 3p       & $\mathrm{^{2} P^{o}_{1/2}}$ \\
\Ion{Si}{4} & 1533&219 & 4d       & $\mathrm{^{2} D^{~}_{5/2}}$ &--& 6f       & $\mathrm{^{2} F^{o}_{7/2}}$ \\
\Ion{Si}{4} & 1533&219 & 4d       & $\mathrm{^{2} D^{~}_{5/2}}$ &--& 6f       & $\mathrm{^{2} F^{o}_{5/2}}$ \\
\Ion{Si}{4} & 1533&222 & 4d       & $\mathrm{^{2} D^{~}_{3/2}}$ &--& 6f       & $\mathrm{^{2} F^{o}_{5/2}}$ \\
\Ion{Si}{4} & 1722&526 & 3d       & $\mathrm{^{2} D^{~}_{5/2}}$ &--& 4p       & $\mathrm{^{2} P^{o}_{3/2}}$ \\
\Ion{Si}{4} & 1722&562 & 3d       & $\mathrm{^{2} D^{~}_{3/2}}$ &--& 4p       & $\mathrm{^{2} P^{o}_{3/2}}$ \\
\Ion{Si}{4} & 1727&376 & 3d       & $\mathrm{^{2} D^{~}_{3/2}}$ &--& 4p       & $\mathrm{^{2} P^{o}_{1/2}}$ \\
\Ion{Si}{5} & 1235&453 & 3s       & $\mathrm{^{3} P^{o}_{2}  }$ &--& 3p       & $\mathrm{^{3} D^{~}_{2}  }$ \\
\Ion{Si}{5} & 1251&390 & 3s       & $\mathrm{^{3} P^{o}_{2}  }$ &--& 3p       & $\mathrm{^{3} D^{~}_{3}  }$ \\
\Ion{Si}{5} & 1276&008 & 3s       & $\mathrm{^{3} P^{o}_{1}  }$ &--& 3p       & $\mathrm{^{3} D^{~}_{2}  }$ \\
\Ion{Si}{5} & 1285&458 & 3s       & $\mathrm{^{3} P^{o}_{0}  }$ &--& 3p       & $\mathrm{^{3} D^{~}_{1}  }$ \\
\Ion{Si}{5} & 1319&600 & 3s       & $\mathrm{^{1} P^{o}_{1}  }$ &--& 3p       & $\mathrm{^{1} D^{~}_{2}  }$ \\
\noalign{\smallskip}
\Ion{Ar}{6} & 1283&934 & 3p$^{2}$ & $\mathrm{^{2} P^{~}_{1/2}}$ &--& 3p$^{3}$ & $\mathrm{^{2} D^{o}_{3/2}}$ \\
\Ion{Ar}{6} & 1303&864 & 3p$^{2}$ & $\mathrm{^{2} P^{~}_{3/2}}$ &--& 3p$^{3}$ & $\mathrm{^{2} D^{o}_{3/2}}$ \\
\Ion{Ar}{6} & 1307&349 & 3p$^{2}$ & $\mathrm{^{2} P^{~}_{3/2}}$ &--& 3p$^{3}$ & $\mathrm{^{2} D^{o}_{5/2}}$ \\
\noalign{\smallskip}
\hline
\end{tabular}
\end{table}

\begin{figure}[ht]
  \resizebox{\hsize}{!}{\includegraphics{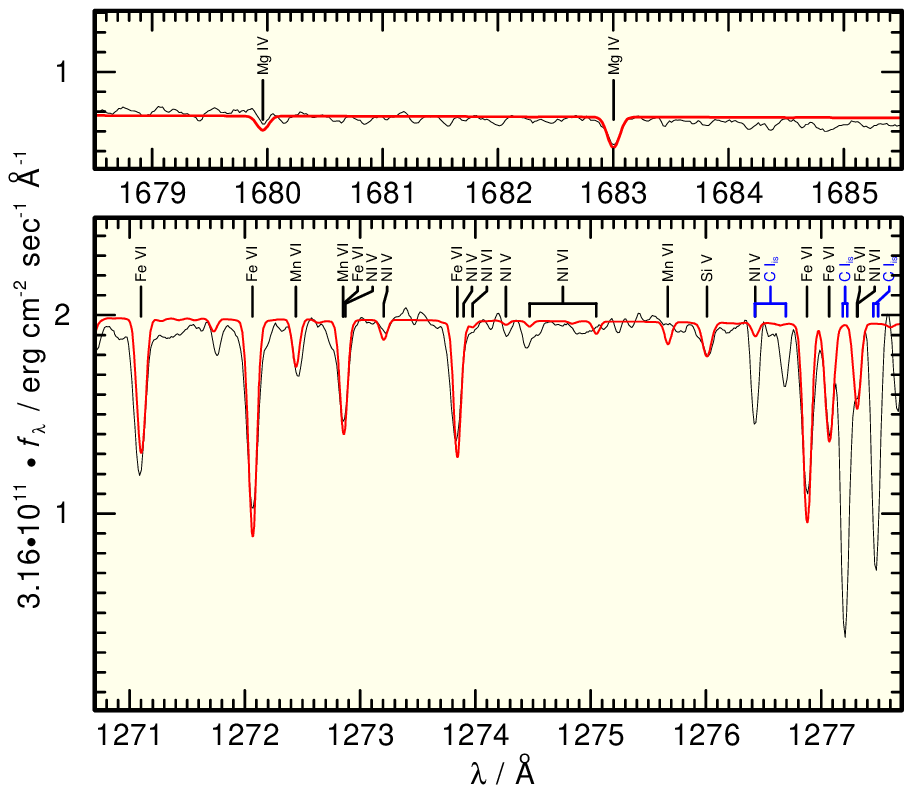}}
  \caption[]{Sections of the STIS spectrum of \lsv\ around \Ionw{Mg}{4}{1683.0} (top) and \Ionw{Si}{5}{1276.0} (bottom).
             Identified lines are marked with their ion's name.
             ``is'' denotes interstellar lines.
            }
  \label{fig:mgsiident}
\end{figure}

Moreover, we could identify \Ion{Si}{4} and \Ion{Si}{5} lines \sT{tab:mgsiar}, which provide
a new ionization equilibrium to evaluate for the \Teff\ determination.
An example for a \Ion{Si}{5} line is shown in \ab{fig:mgsiident}.

We have also identified \Ionw{Ar}{6}{1303.86} as an isolated line in the STIS spectrum \sT{tab:mgsiar} 
of \lsv\ \sA{fig:arident}.
To our best knowledge, this is the first time that \Ion{Ar}{6} has been detected in the photosphere of any star.

\begin{figure}[ht]
  \resizebox{\hsize}{!}{\includegraphics{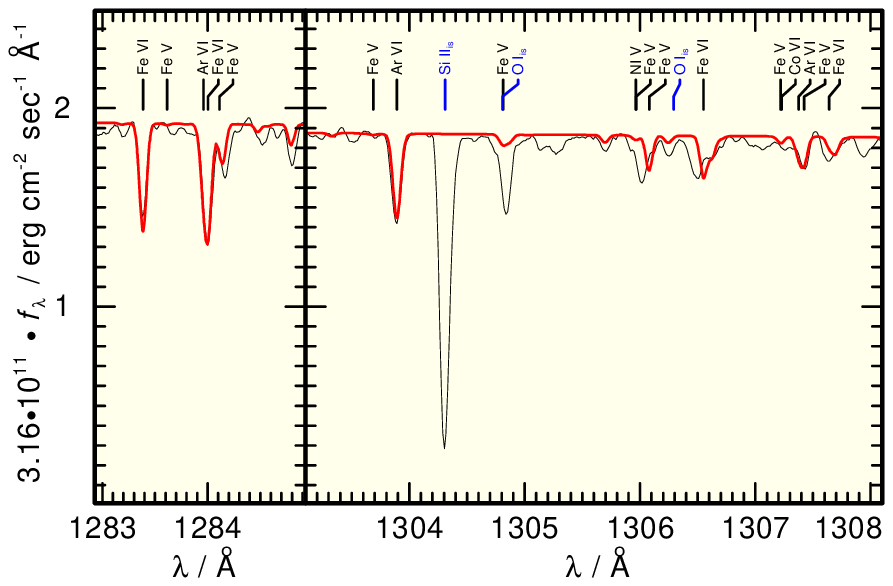}}
  \caption[]{Sections of the STIS spectrum of \lsv\ around \Ionww{Ar}{6}{1283.93, 1303.86, 1307.35}.
            }
  \label{fig:arident}
\end{figure}

Recently, \citet*{wrk2005} have identified
\Ionww{F}{5}{1082.31, 1087.82, 1088.39}
and
\Ionw{F}{6}{1139.50}
in FUSE observations of hot central stars of planetary nebulae (CSPN).
We have inspected the spectrum of \lsv\ but we cannot identify these lines unambiguously \sA{fig:fident}.
However, while a 10$\times$ solar F abundance is definitely too much, these F lines can be hidden in the
spectrum at a solar F abundance. This is fully consistent with the result of \citet{wrk2005} that 
H-rich CSPN have approximately solar F abundances. For our calculations, we adopt a solar F abundance.
  
\begin{figure}[ht]
  \resizebox{\hsize}{!}{\includegraphics{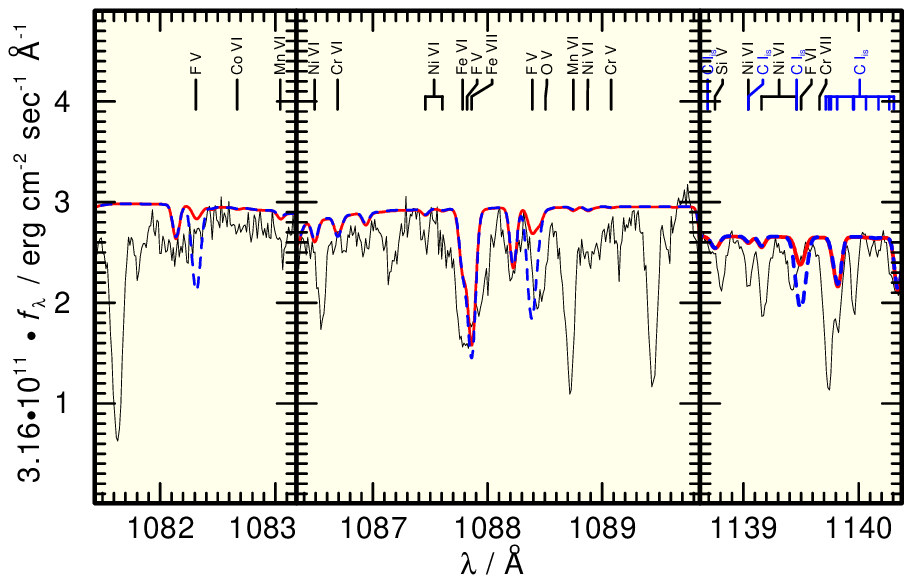}}
  \caption[]{Sections of the FUSE spectrum of \lsv\ around \Ionww{F}{5}{1082.31, 1087.82, 1088.39}
             and \Ionw{F}{6}{1139.50} compared with synthetic spectra with solar (full line) and
             10$\times$ solar F abundance (dashed line).
            }
  \label{fig:fident}
\end{figure}

In addition, lines of H, He, C, N, O, P, S, Cr, Mn, Fe, Co, and Ni have been identified as well.

\begin{figure}[ht]
  \resizebox{\hsize}{!}{\includegraphics{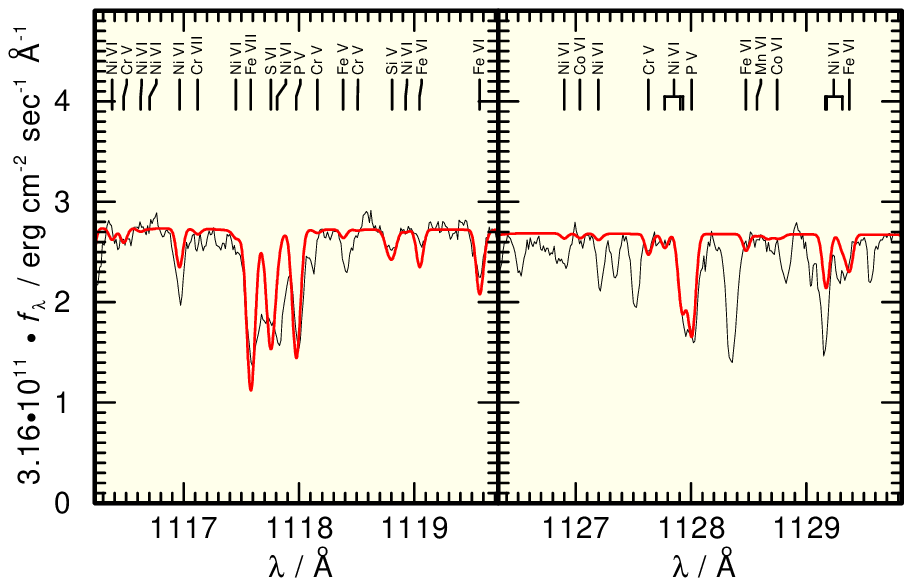}}
  \caption[]{Sections of the FUSE spectrum of \lsv\ around the \Ionww{P}{5}{1118.0, 1128.1} resonance doublet
             compared with our final model.
             Several lines in these sections are not identified.
            }
  \label{fig:pident}
\end{figure}

\begin{figure}[ht]
  \resizebox{\hsize}{!}{\includegraphics{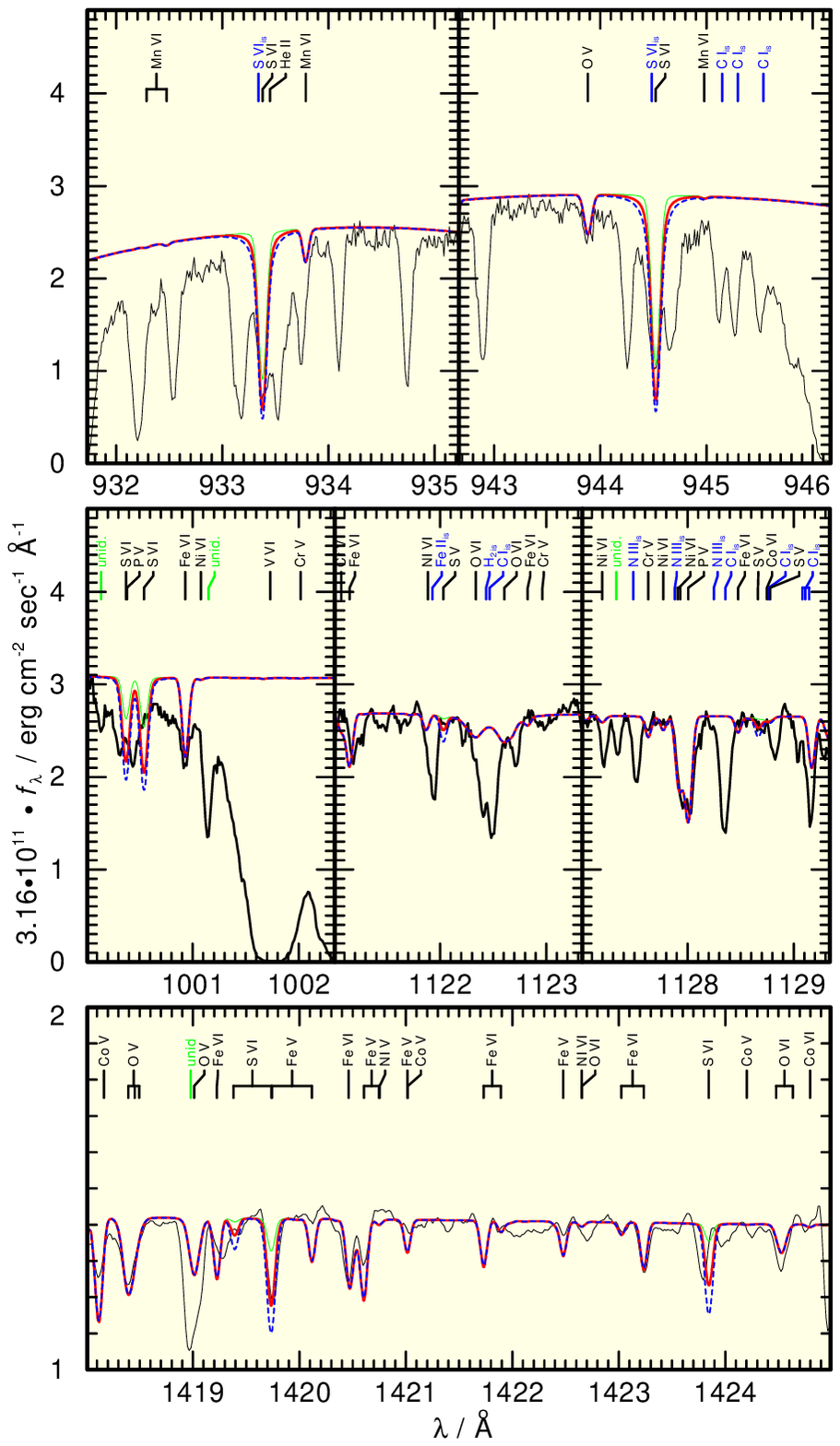}}
  \caption[]{Sections of the FUSE and STIS spectra of \lsv\ around 
             \Ionww{S}{6}{933.4, 944.5} (top),
             \Ionww{S}{5}{1122, 1129} and 
             \Ionw{S}{6}{1000} (middle), and
             \Ionww{S}{6}{1419.38, 1419.74, 1423.85} (bottom).
             The synthetic spectra are calculated from models with solar (dashed line), 0.5$\times$ solar
             (thick line) and 0.1$\times$ solar (thin line) sulfur abundance.
             The strong unmarked absorption features in the top panel stem from interstellar \Ion{H}{1} and H$_2$.
             }
  \label{fig:sident}
\end{figure}

\begin{figure}[ht]
  \resizebox{\hsize}{!}{\includegraphics{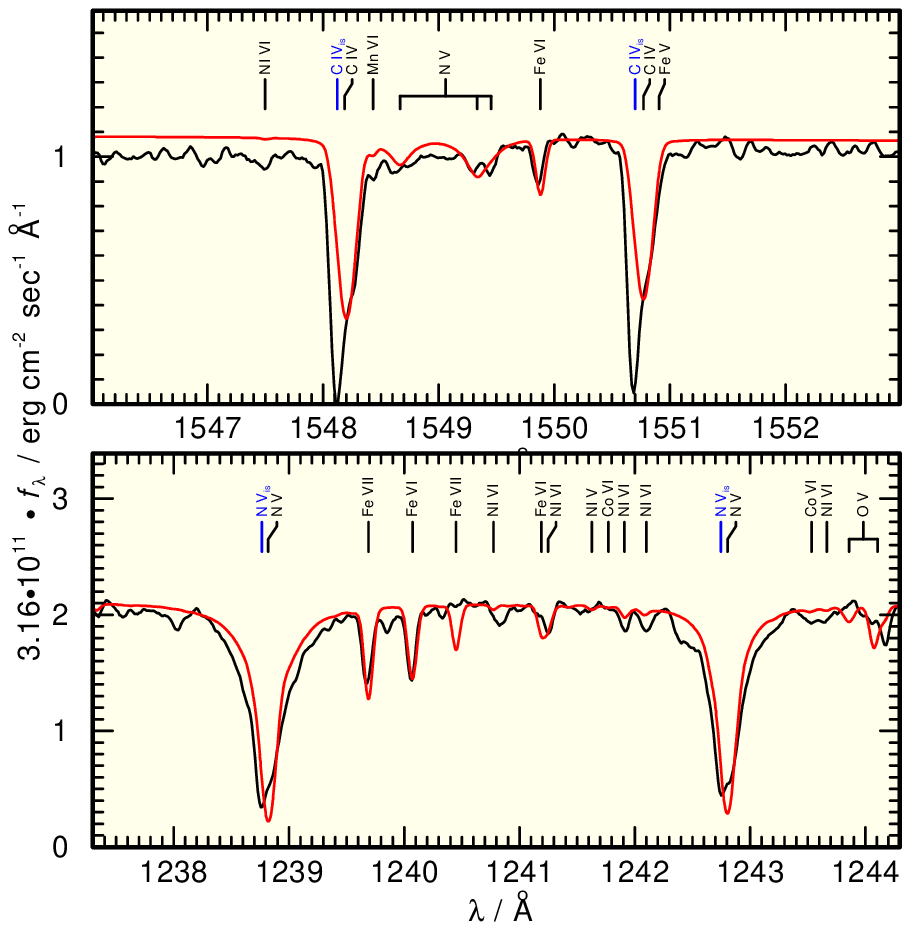}}
  \caption[]{Sections of the STIS spectrum of \lsv\ around
             \Ionww{C}{4}{1548.2, 1550.8}  and
             \Ionww{N}{5}{1238.8, 1242.8}
             compared with our final model.
             The \Ionw{Fe}{7}{1240.4} line is obviously too strong, 
             probably due to an error (f-value too high) in Kurucz's line list \citep{k1996}.
            }
  \label{fig:resonance}
\end{figure}

\begin{figure}[ht]
  \resizebox{\hsize}{!}{\includegraphics{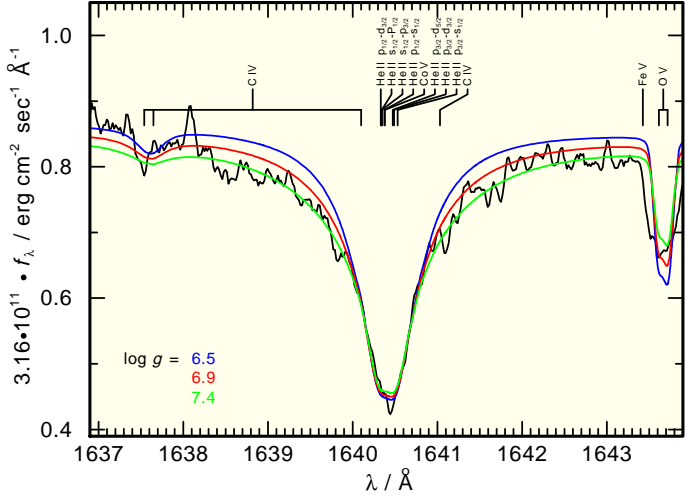}}
  \caption[]{Section of the STIS spectrum of \lsv\ around \Ionww{He}{2}{1640.32 - 1640.53} compared with
             theoretical line profiles calculated from models with \Teffw{95} and \loggw{6.5, 6.9, 7.3}.
             Note that the width of the fine-structure splitting amounts to $\approx 0.22\,\mathrm{\AA}$ 
             and thus, has to be considered at STIS's resolution here. For the \Ion{He}{2} line broadening, we use
             tables by \citet{sb1989}.
             }
  \label{fig:heii1640}
\end{figure}

\begin{figure}[ht]
  \resizebox{\hsize}{!}{\includegraphics{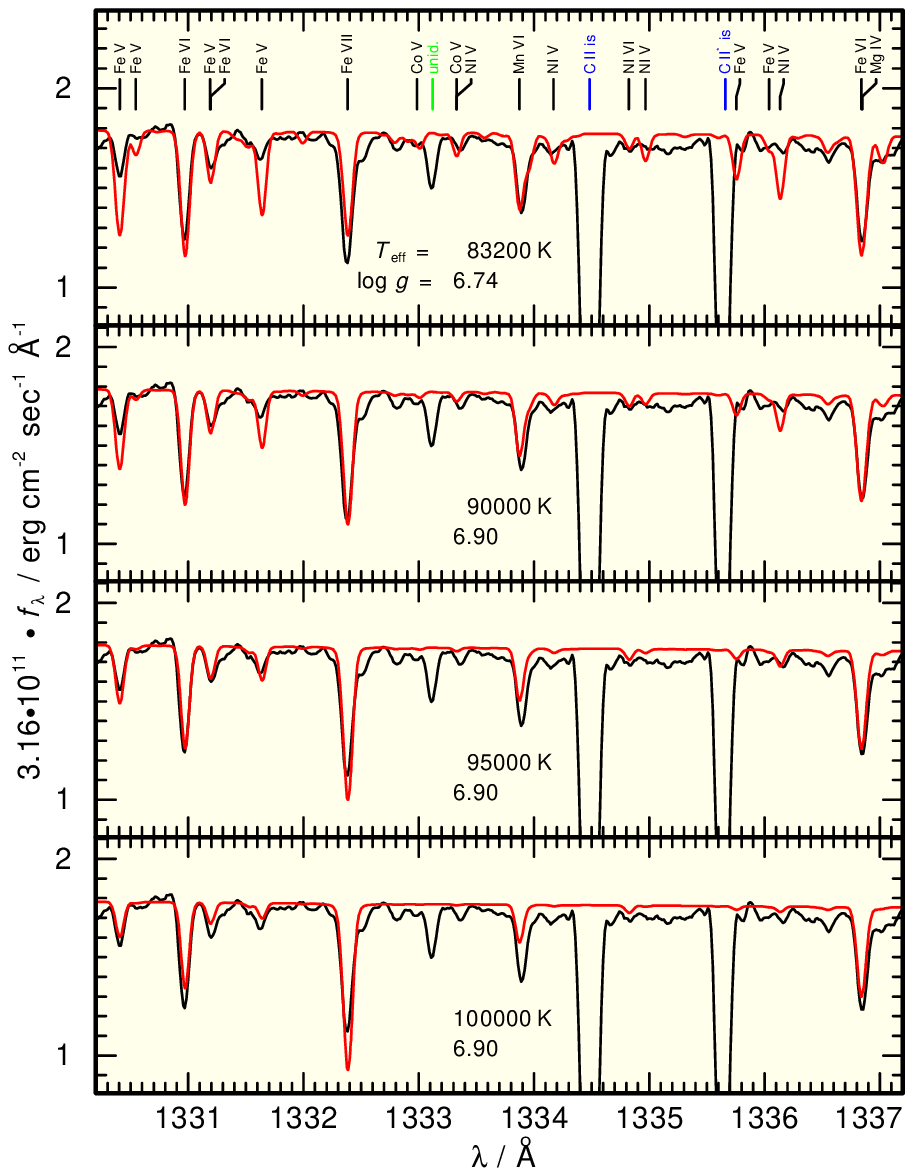}}
  \caption[]{Section of the STIS spectrum of \lsv\ compared to models with different \Teff\ and \logg.
             The \Ion{Fe}{5} and \Ion{Fe}{7} lines appear strongly dependent on changes of \Teff\
             while \Ion{Fe}{6} is almost independent in the relevant \Teff\ range.
             \Ion{Fe}{5}, \Ion{Fe}{6}, and \Ion{Fe}{7} are reproduced simultaneously at \Teffw{95}.
             Note that the \Teff\ determination by \citet*{n1999} based on H+He composed models
             resulted in a much too low \Teff\ (top panel). 
             ``unid\@.'' denotes a unidentified line.
            }
  \label{fig:tefffe}
\end{figure}

\begin{figure}[ht]
  \resizebox{\hsize}{!}{\includegraphics{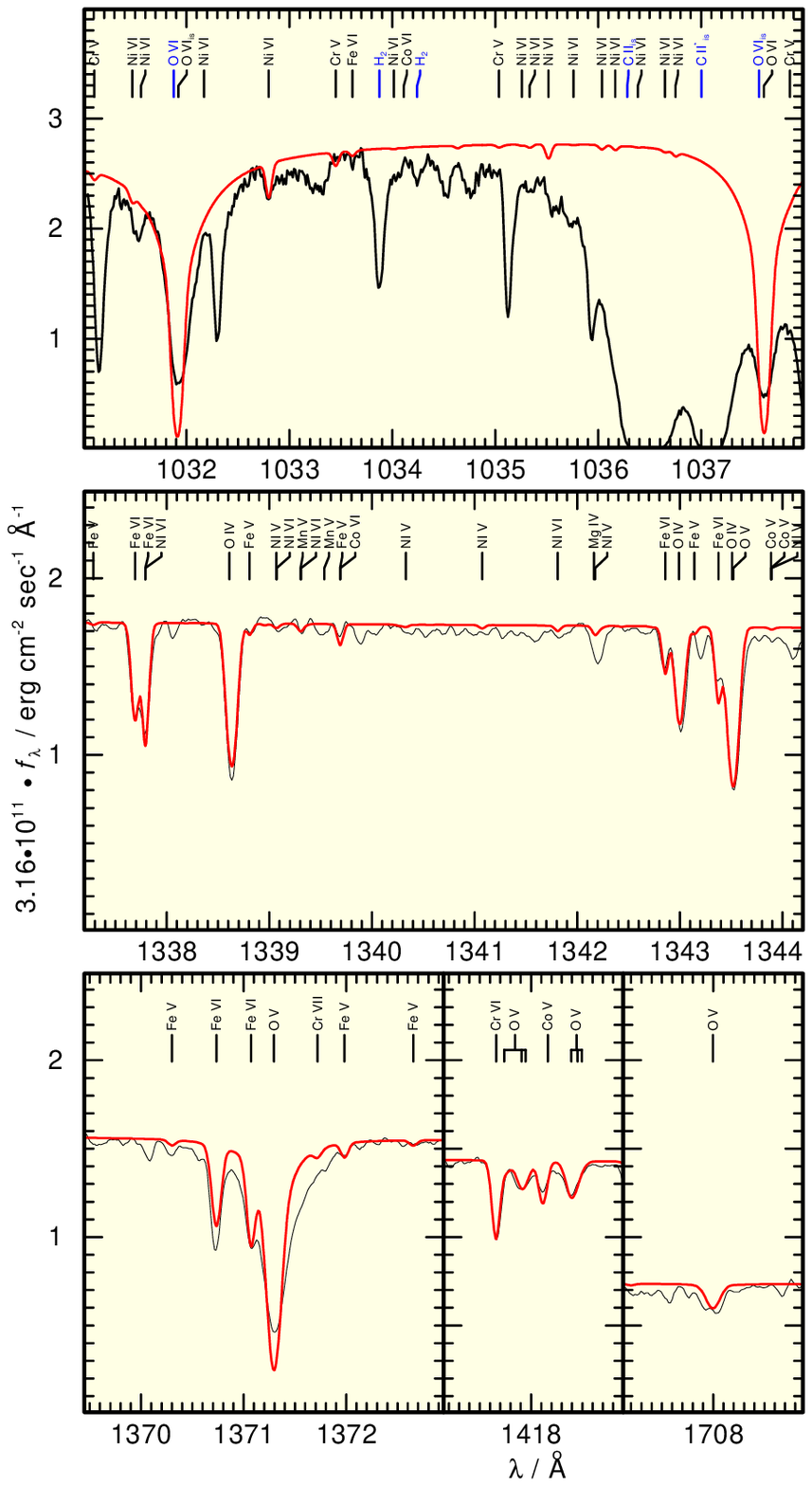}}
  \caption[]{Sections of the FUSE and STIS spectra of \lsv\ around 
             \Ionww{O}{6}{1031.9, 1037.6} (top),
             \Ionww{O}{4}{1338.6, 1343.0, 1343.5} (middle),
             \Ionw{O}{5}{1371.3} (bottom, left),
             \Ionww{O}{5}{1417.7. 1417.9, 1418.4} (bottom, middle), and
             \Ionw{O}{5}{1708.0} (bottom, right)
             compared with our final model.
            }
  \label{fig:teffo}
\end{figure}

\begin{figure}[ht]
  \resizebox{\hsize}{!}{\includegraphics{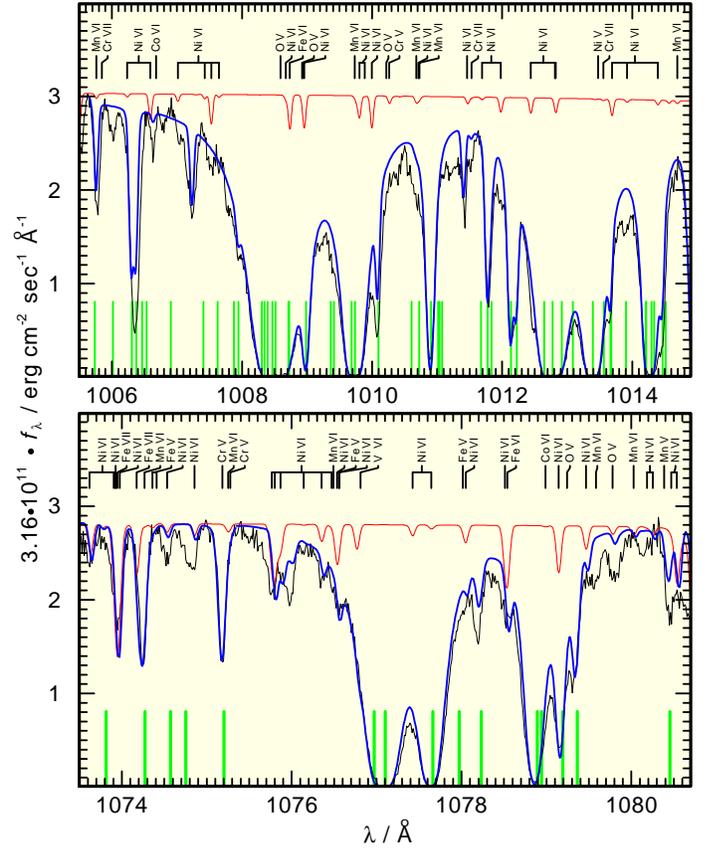}}
  \caption[]{Sections of the FUSE spectrum of \lsv\ compared 
             with our best model of the ISM absorption (top panel, thick line) and
             with the combined ISM + model-atmosphere spectrum (bottom panel, thick line).
             The thin lines are the pure model-atmosphere spectrum.
             Most of the interstellar absorption is due to H$_2$ (the line positions
             are marked by vertical bars at the bottom of the panels).
             The synthetic spectra are normalized to match the local continuum. 
             Note that it is possible to identify a few isolated lines, 
             e.g\@. \Ionw{Fe}{7}{1073.9}, which are suitable for spectral analysis.
            }
  \label{fig:owens}
\end{figure}

The complete FUSE and STIS spectra compared to the synthetic spectra calculated from our final model with identification marks 
as well as a table with the wavelengths of all identified lines are available online.
In Figs\@. 
\ref{fig:nhi}, 
\ref{fig:fident} --
\ref{fig:sident},
\ref{fig:teffo}, and
\ref{fig:owens},
we show some details of the
FUSE spectrum. 
In Figs\@. 
\ref{fig:nhi},
\ref{fig:mgsiident},
\ref{fig:arident},
\ref{fig:sident} --
\ref{fig:heii1640},
\ref{fig:tefffe}, and
\ref{fig:teffo},
we show some details of the
STIS spectrum. 
These figures show representative samples of absorption by various photospheric constituents; discussion of
the individual species is given in subsequent sections.

The observed spectra have been shifted to the rest wavelength of the photospheric lines.

\subsection{Analysis}
\label{subsect:analysis}
 
In the first part of this analysis, we will mainly concentrate on the STIS spectrum in which 
we find many isolated lines of many
species and their ions. Since the number of species is large, it is impossible to
calculate extended model-atmosphere grids on a reasonable time scale. 
Therefore, we pursued the following strategy.
We start our analysis with model atmospheres based on parameters \sK{sect:introduction} of \citet{tea2005}. We
include Mg and Ca -- Ni in addition (solar abundance ratios). 
In a first step, we will then re-adjust \Teff\ precisely and check \logg\ \sK{subsubsect:teff}. 
Subsequently, we will fine-tune the C, N, and O abundances in order to improve the
fit to the STIS observation \sK{subsubsect:abundances}.
The abundances of Mg, Si, P, S, Ar, and Ca -- Ni are then fine-tuned.
During this whole process, our results for \Teff\ and \logg\ are continuously checked for consistency.

\subsubsection{Metal abundances}
\label{subsubsect:abundances}

\begin{figure}[ht]
  \resizebox{\hsize}{!}{\includegraphics{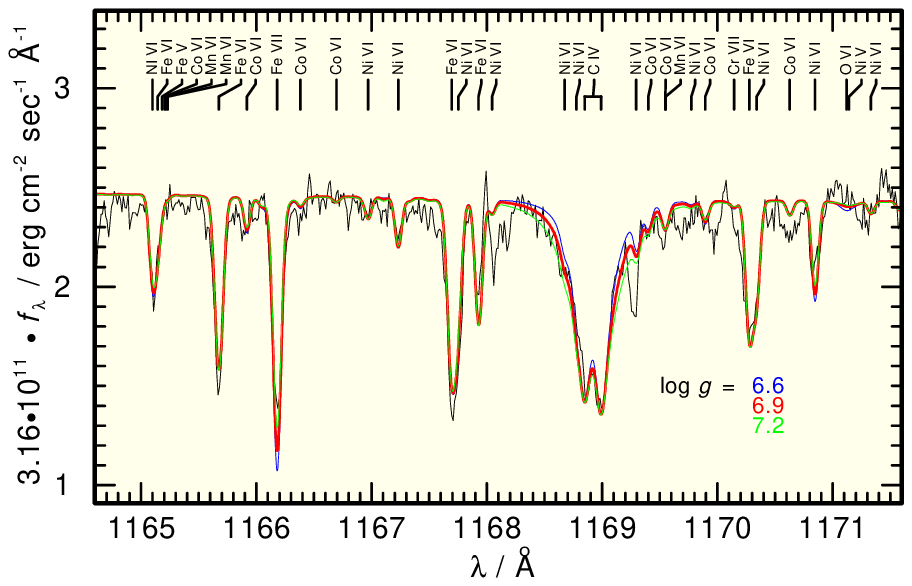}}
  \caption[]{Sections of the FUSE spectrum of \lsv\ around 
             \Ionww{C}{4}{1168.85, 1168.99}
             compared with synthetic spectra calculated from models with different \logg.
             While at \loggw{6.6} the line wings of \Ionww{C}{4}{1168.85, 1168.99} are
             too narrow, they appear too broad at \loggw{7.2}. The observation is best matched
             at \loggw{6.9}.
            }
  \label{fig:1164-1171}
\end{figure}

The abundances of C, N, O, Mg, and Si have been determined by detailed line-profile fits 
based on the FUSE and STIS observations. 
Examples are shown in Figs\@. \ref{fig:mgsiident}, \ref{fig:1164-1171}, and \ref{fig:teffo}.

In the FUSE spectrum of \lsv, \Ionww{P}{5}{1117, 1128} are identified \sA{fig:pident}. 
These lines are well reproduced at a 0.5$\times$ solar abundance.

We have identified 
\Ionww{S}{6}{933, 944} in the FUSE and 
\Ionww{S}{6}{1419.38, 1419.74, 1423.85} in the STIS spectra \sA{fig:sident}.
The line cores of the \Ion{S}{6} resonance doublet appear too deep to match the observation
and are not well-suited for an abundance determination.
The reason might be the uncertain continuum level in the FUSE wavelength range due to the
strong interstellar absorption and the unsufficient line-broadening tables. Though we use data
of \citet{ds1993}, these need to be extrapolated to the temperatures and densities at
the line-forming regions especially of the line cores (which form in the outer parts of the
atmosphere) and are, thus, not very accurate.

However, 
\Ionww{S}{6}{1419.38, 1419.74, 1423.85} \sA{fig:sident} are well reproduced at a 0.5$\times$ solar abundance.
\Ionw{S}{6}{1117.76} \sA{fig:pident} is part of a strong blend but is in reasonable agreement
with the observation.
\Ionww{S}{5}{1122, 1129, 1134} as well as 
\Ionww{S}{6}{976, 1000, 1018} are not identified 
but these are too strong at solar abundance.

We find discrepancies similar to those with the \Ion{S}{6} resonance doublet's line cores (too deep)
also for \Ion{O}{6} \sA{fig:teffo} where we use tables provided by \citet{ds1992a}.
We encounter the same problems in the STIS wavelength region where the continuum is much better defined
with the resonance doublet of \Ion{N}{5} (\ab{fig:resonance}, tables by \citet{ds1992b}). 
In the case of \Ion{C}{4} (\ab{fig:resonance}, tables by \citet{dsb1991}), 
the absorption in the model is too weak; a strong interstellar absorption component may be the explanation.

In the STIS spectrum of \lsv, we could identify iron-group lines of Cr, Mn, Fe, Co, and Ni. For these we determined
the abundances \sA{fig:abundances} with an accuracy of 0.3\,dex from line profile fits.
No trace of Ca, V, Sc, and Ti was found, neither in the STIS nor in the FUSE spectrum.
We did some test calculations in order to find upper abundance limits, i.e\@. at what abundances do lines of
these species appear too strongly to be in hidden in the observation. These limits are about 20 -- 50 times
solar. For our model calculations, however, we assume that the Ca, V, Sc, and Ti 
have a solar-relative abundance pattern with the other iron-group elements and adopt a $5\times$ solar
abundance for them.

\subsubsection{Surface gravity}
\label{subsubsect:g}

\citet{tea2005} calculated a spectroscopic distance of $d=240\pm 36\,\mathrm{pc}$ 
(same method as described in \se{subsect:MdL}). 
This is not in agreement with the result of \citet{hea2007} who found $d=128.9^{+5.7}_{-5.3}\,\mathrm{pc}$
from the trigonometric parallax. 
The spectroscopic distance is strongly dependent on \logg\ (see \eq{Eq:d}) which is determined from detailed line profile fits.
From our experience, we know that in the relevant \logg\ range (around 7), the typical error range is about 0.5\,dex --
resulting in a large error in $d$.
Since we have a direct measurement of the distance \citep{hea2007}, this adds a strong constraint \sK{subsect:MdL}.
From \eq{Eq:d}, we can estimate that we need a higher \logg\ (of about 7.4) than found in previous analyses and thus,
we extend our model-atmosphere grid to higher \logg\ and higher \Teff.
However, we find that a higher \logg\ cannot be compensated by a change of \Teff\ within the error limits
derived from fits to the hydrogen Balmer lines.
A detailed comparison of the \Ionw{He}{2}{1640} 2-3 multiplet with the STIS observation is shown in \ab{fig:heii1640}.
\Ionw{He}{2}{1640} is sensitive to both the He abundance as well as \logg. E.g\@. a factor of
0.5/2.0 in the He abundance can be approximately compensated by +0.5/$-$0.5 in \logg. 
However, from this line we derive \loggw{6.9\pm 0.3} and [He] $= 0.1\pm 0.3\,\mathrm{dex}$.
At \loggw{6.9}, e.g\@. the \Ionw{C}{4}{1169} doublet is well reproduced \sA{fig:1164-1171}.

The determination of \logg\ is a crucial issue of this work and thus, 
we tried to use the hydrogen Lyman lines for this purpose.
Unfortunately, neither L\,$\alpha$ which is in the STIS wavelength range
nor the higher lines of the Lyman series are suited:
while the complete L\,$\alpha$ line profile is dominated by absorption of interstellar \Ion{H}{1} \sA{fig:nhi},
the strong interstellar absorption in the FUSE wavelength range at $\lambda \sla 1100\,\mathrm{\AA}$ strongly
hampers, e.g., the determination of the local continuum.
Thus, we have a look at the hydrogen Balmer lines 
because their line broadening is well known and detailed \Ion{H}{1} line-broadening
tables are available \citep{l1997}. In \ab{fig:logg_Hb} we show H\,$\beta$ as an example.
The determination of \loggw{6.9\pm 0.2} from the UV wavelength range is confirmed within even narrower error limits.
Plots of the model spectrum computed with our final adopted parameters
are shown with the observed Balmer line profiles in \ab{fig:blp2006}.

\begin{figure}[ht]
  \resizebox{\hsize}{!}{\includegraphics{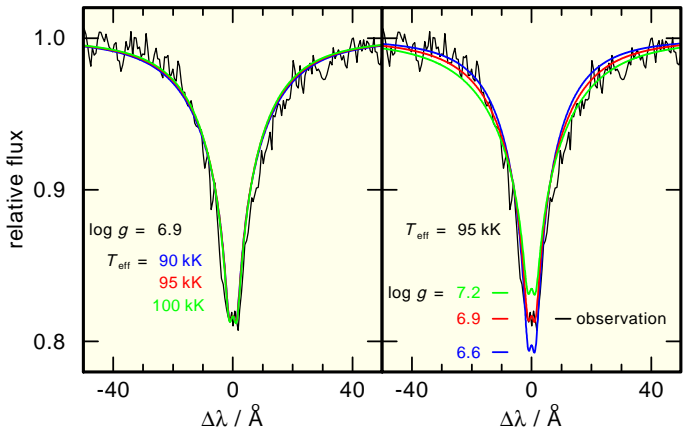}}
  \caption[]{Dependency of H\,$\beta$ on \Teff\ (left) and \logg\ (right).
             A section of the optical spectrum of \lsv\ (taken in Oct 1990 with the TWIN spectrograph
             attached to the 3.5\,m telescope at the Calar Alto observatory) 
             around H\,$\beta$ compared with
             theoretical line profiles calculated from models with \Teffw{90, 95, 100} and \loggw{6.5, 6.9, 7.3}.
             In this parameter range, H\,$\beta$ appears not very sensitive on \Teff\ but very sensitive
             on \logg\ in the line center.
             At \loggw{6.9}, the central depression is perfectly matched and the observed line profile is well
             reproduced.
             }
  \label{fig:logg_Hb}
\end{figure}

\subsubsection{Effective temperature}
\label{subsubsect:teff}

An ionization equilibrium is a very sensitive indicator for \Teff. 
The evaluation of ionization equilibria of many species with many lines increases the accuracy \citep[e.g\@.][]{r1993} further. 
As a prerequisite, lines of successive ions of a species have to be identified. 
In the STIS spectrum of \lsv, we found lines of
\Ion{N}{4}  -- \Ion{N}{5},
\Ion{O}{4}  -- \Ion{O}{5},
\Ion{Si}{4} -- \Ion{Si}{5},
\Ion{Fe}{5} -- \Ion{Fe}{7}, and
\Ion{Ni}{5} -- \Ion{Ni}{7}
which are suitable for this purpose. 
In addition, the FUSE spectrum provides lines of
\Ion{N}{4}, 
\Ion{O}{6}, 
\Ion{Si}{4},
\Ion{Fe}{6} -- \Ion{Fe}{7}, and
\Ion{Ni}{6} -- \Ion{Ni}{7}.

With our synthetic spectra, we are able to reproduce all O lines but \Ionw{O}{5}{1371}. 
The reason is unknown, however, a possible reason is the approximate formula used for
the quadratic Stark effect. Thus, detailed line-broadening data for this line is highly desirable.

In \ab{fig:tefffe} we show the dependency of the \Ion{Fe}{5} -- \Ion{Fe}{7} equilibrium on \Teff.
We can model \Ion{Fe}{5} -- \Ion{Fe}{7} lines simultaneously at \Teffw{95}.
\ab{fig:teffo} demonstrates that other ionization equilibria are also well reproduced at this \Teff. 
For other species, the reader may have a look at the complete FUSE and STIS spectra \sK{subsect:lines}.

\begin{figure}[ht]
  \resizebox{\hsize}{!}{\includegraphics{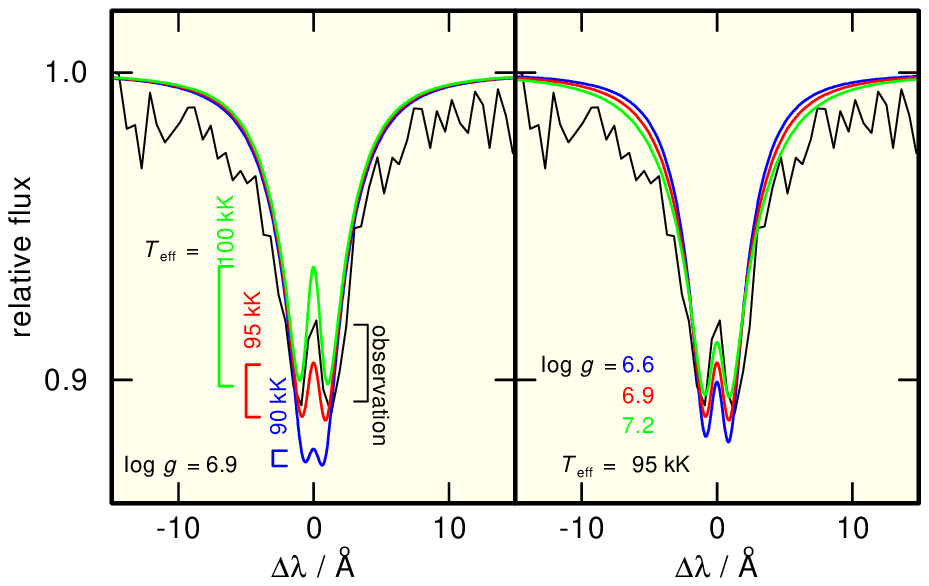}}
  \caption[]{Dependency of \Ionw{He}{2}{4686} on \Teff\ (left) and \logg\ (right).
             A section of the optical spectrum of \lsv\ around \Ionw{He}{2}{4686} compared with
             theoretical line profiles calculated from models with \Teffw{90, 95, 100} and \loggw{6.5, 6.9, 7.3}.
             In this parameter range, \Ionw{He}{2}{4686} appears not very sensitive on \logg\ but very sensitive
             on \Teff\ in the line center.
             The brackets indicate the ``strength'' of the emission reversal in the line core.
             }
  \label{fig:HeII4686}
\end{figure}

\begin{figure}[ht]
  \resizebox{\hsize}{!}{\includegraphics{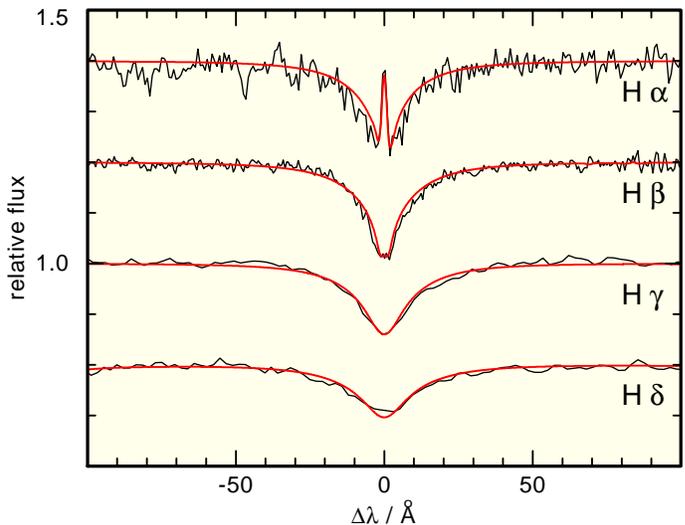}}
  \caption[]{Synthetic Balmer-line spectra of our final model (\Teffw{95}, \loggw{6.9}) compared with the observation
             \citep[cf.][]{n1999}. 
            }
  \label{fig:blp2006}
\end{figure}

Additional constraints for \Teff\ are found in the optical wavelength range, where the observed
H\,$\alpha$ and \Ionw{He}{2}{4686} line profiles (Figs\@. \ref{fig:blp2006}, \ref{fig:HeII4686}) 
show emission reversals in their line cores. These are reproduced by our models. 
Such emission cores can be used as a measure for \Teff\ as well \citep*{rkw1996}.
While the H\,$\alpha$ emission core changes only little in the relevant \Teff\ range, significant
changes are visible in \Ionw{He}{2}{4686} \sA{fig:HeII4686}. From this line, \Teffw{97} can be
determined. However, precise \Teff\ determinations from emission reversal profiles require
higher spectral resolution than the 1.5\,\AA\ achieved in the present spectrum.

We finally adopt \Teffw{95}.
From the comparison of synthetic spectra from models within \Teffw{90 - 100}, 
we estimate an error range in \Teff\ of 2\,kK.

\subsection{Mass, distance, and luminosity}
\label{subsect:MdL}

In \ab{fig:ltlg} we compare the position of \lsv\ to evolutionary tracks
in the $\log \Teff - \logg$ plane. 
We can interpolate a mass of $M = 0.550^{+0.020}_{-0.015}\,\mathrm{M_\odot}$ and
a luminosity of $\log L/\mathrm{L_\odot} = 2.2\pm 0.2$ from the evolutionary tracks
of \citet{s1983} and \citet{bs1990}.

\begin{figure}[ht]
  \resizebox{\hsize}{!}{\includegraphics{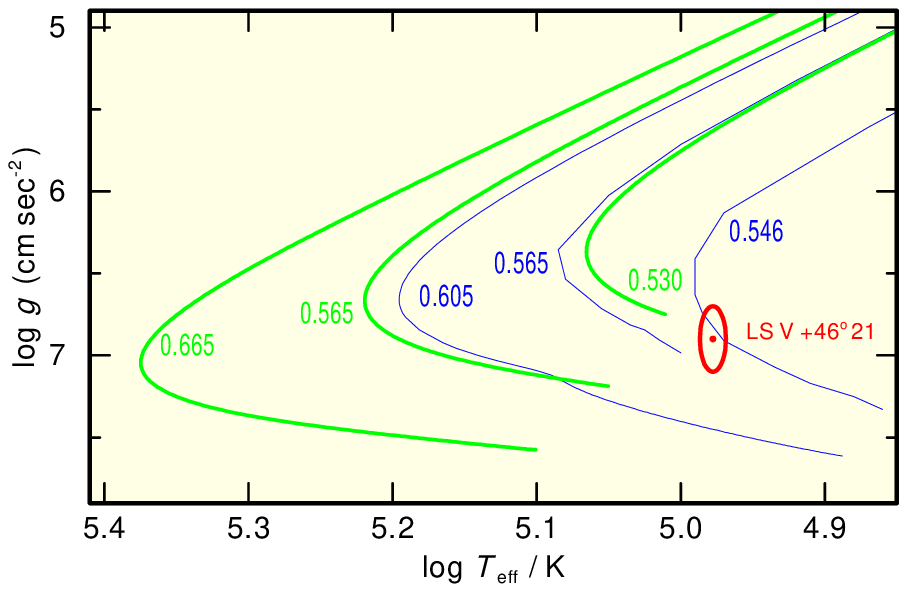}}
  \caption[]{Position (error ellipse) of \lsv\ in the $\log \Teff\ - \logg$ plane compared with evolutionary tracks
             of hydrogen-burning post-AGB stars \citep[][thin lines]{s1983, bs1990}.
             The labels give the mass of the remnant in $\mathrm M_\odot$.
             Note that a comparison with the new tracks by \citet[][thick lines]{aea2005} yields a lower 
             remnant mass.  
            }
  \label{fig:ltlg}
\end{figure}

The spectroscopic distance of \lsv\ is calculated following the flux calibration of \citet{hea1984}, 

\begin{equation}
\label{Eq:fv}
f_{\rm V} = 3.58\times 10^{-9}\times 10^{{\rm -0.4}m_\mathrm{V_0}}\, \mathrm{erg\,cm^{-2}\,sec^{-1}\,\AA\hbox{}^{-1}}
\end{equation}

\noindent
with 
$m_\mathrm{V_0} = m_\mathrm{V} - 2.175 c$, 
$c = 1.47 E_\mathrm{B-V} = 0.096^{+0.014}_{-0.022}$,
$m_\mathrm{V} = 12.67\pm 0.02$, 
\loggw{6.9\pm 0.2}, and 
$M = 0.530^{+0.020}_{-0.015} {\rm M_\odot}$, 
the distance is derived from

\begin{equation}
\label{Eq:d}
d = 7.11\times 10^{4} \sqrt{H_\nu\times M\times 10^{0.4m_\mathrm{V_0}-\log g}}~{\rm pc}\,.
\end{equation}

\noindent
With the Eddington flux at $\lambda_\mathrm{eff} = 5454\,\mathrm{\AA}$ of our final model atmosphere,
$H_\nu = (1.49\pm 0.03) \times 10^{-3} \mathrm{erg\,cm^{-2}\,sec^{-1}\,Hz^{-1}}$,
we derive a distance of $d= 224^{+46}_{-58}\,\mathrm{pc}$. 
This is not in agreement with the distance ($d=128.9\,\mathrm{pc}$) of \citet{hea2007}.
On the basis of the spectral analysis a value of \loggw{7.4} which would be needed to reach
agreement can be excluded.
\sK{subsubsect:g}.


                                         


\section{The FUSE spectrum and the ISM absorption model}
\label{sect:owens}

Along the line of sight to \lsv\ we detect interstellar absorption by
\Ion{H}{1}, 
\Ion{D}{1}, 
\Ion{C}{1}, 
\Ion{C}{1}$^*$, 
\Ion{C}{1}$^{**}$, 
\Ion{C}{2}, 
\Ion{C}{2}$^*$, 
\Ion{N}{1}, 
\Ion{N}{2}, 
\Ion{O}{1}, 
\Ion{Al}{2},
\Ion{Si}{2}, 
\Ion{Si}{3}, 
\Ion{P}{2}, 
\Ion{S}{1}, 
\Ion{S}{2}, 
\Ion{Cl}{1}, 
\Ion{Ar}{1}, 
\Ion{Fe}{2}, 
\Ion{Fe}{3}, 
H$_2$ ($J$ = 0 -- 9), 
HD ($J$ = 0 -- 1),
and CO. 
Absorption by 
\Ionww{C}{4}{1548, 1550}, 
\Ionww{N}{5}{1238, 1242}, and 
\Ionww{Si}{4}{1393, 1402} 
is also present at the velocity of the interstellar absorption of the species mentioned above 
($v_{\rm helio} \approx 6\,\mathrm{km\,s^{-1}}$), 
in addition to the photospheric absorption at 
$\approx 20\,\mathrm{km\,s^{-1}}$.

At the FUSE and STIS resolutions all the interstellar absorption lines
display only a single absorption component with a common velocity,
$v_{\rm helio} \approx 6\,\mathrm{km\,s^{-1}}$.
However, the detection of several ionization stages for some of
the species (e.g\@. \Ion{C}{1}, \Ion{C}{2}, and \Ion{C}{4}) 
indicate that there must be at least
three ISM components along this line of sight: a cool component traced by
molecular hydrogen and \Ion{C}{1} (amongst other species), a photoionized
component traced by the high-ionization species (\Ion{C}{4}, \Ion{N}{5}, and \Ion{Si}{4}),
and a warm component where the bulk of the species such as \Ion{C}{2}, \Ion{N}{1}, \Ion{O}{1},
\Ion{S}{2}, \Ion{Fe}{2}, etc\@. resides.
The highly ionized species likely reside in the photoionized gas associated
with the PN where optical emission lines of H\,$\alpha$ and [\Ion{O}{3}] 
have been detected \citep{f1981}.

The analysis of the ISM along this line of sight is discussed by Oliveira (2007, in prep.).

With the procedure {\sc OWENS} it is possible to model different interstellar absorption clouds with different chemical 
compositions, radial and turbulent velocities, temperature and column densities for each element included.
We show two examples of the quality of our ISM modeling \sA{fig:owens}.

The ISM fit is generally quite good, but incorporation of the new stellar model into the
fits will help to refine the derived ISM parameters.

\section{The Galactic orbit}
\label{sect:orbit}

We used the STIS spectrum of \lsv\ in order to determine its
radial velocity out of a total of 54 \Ion{Fe}{6} and \Ion{Fe}{7} photospheric lines. 
We obtained 
$v_\mathrm{rad} = +20.6\,\mathrm{km\,s^{-1}}$ with a standard deviation of 
$\sigma_{v_\mathrm{rad}} = \pm 1.5\,\mathrm{km\,s^{-1}}$. 
This is in agreement with $v_\mathrm{rad} = +22.4\pm 3\,\mathrm{km\,s^{-1}}$ measured by \citet{tea2005}.
Measurements from IUE spectra by \citet{tn1992} and \citet{hea1998} (11.9 and 11.1\,km\,s$^{-1}$, respectively)
may be less accurate because the star was possibly not well centered in the aperture (Holberg priv\@. comm.).

\citet{kea2004} have performed their orbit calculations with the radial velocity given by \citet{tn1992} which is about
a factor of two lower than measured from our STIS spectra. 
Thus, we have re-calculated the orbit \sA{fig:orbit}.
We follow \citet*{pea2003} and
use the Galactic potential of \citet*{fea1996} which includes a dark halo, bulge and stellar components, 
as well as three disks.
The equations of motion are integrated using a fourth order Runge-Kutta scheme.
To compute the tangential velocity we have used as a reference the proper motion of 
\citet{kea2004}. 
Using the parallactic distance (128.9\,pc) of \citet{hea2007}, we find that
\lsv\ is a thin-disk object and its orbit extends about 
$\pm 0.20\,\mathrm{kpc}$ perpendicular to the Galactic plane
at a distance interval of $8.02\,\mathrm{kpc} < \rho < 8.85\,\mathrm{kpc}$ from the Galactic center \sA{fig:orbit}.
With our spectroscopic distance (224\,pc), 
there is not much difference: the orbit is now confined between $\pm 0.25\,\mathrm{kpc}$ from the Galactic plane,
whereas it is lying at a distance $7.50\,\mathrm{kpc} < \rho < 8.90\,\mathrm{kpc}$ from the center.

We have verified that this conserves total energy to better than $10^{-11}$, 
and the $z$ component of angular momentum to better than $10^{-10}$ for \lsv.

\begin{figure}[ht]
  \resizebox{\hsize}{!}{\includegraphics{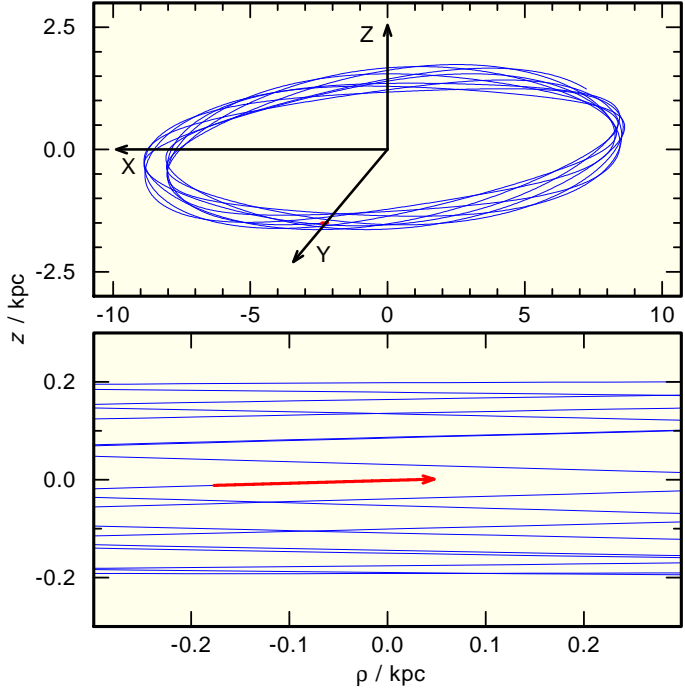}}
  \caption[]{The Galactic orbit of \lsv\ in the last 2\,Gyrs in Galacto-centric coordinates (top).
             At the bottom, a magnification around the present position of \lsv\ is shown.
             During its PN phase ($\approx$460\,000 years), 
             \lsv\ has made its way along the (very small) thicker part of the track.
            }
  \label{fig:orbit}
\end{figure}

The location of \lsv\ is presently about 1\,pc above the Galactic plane \sA{fig:orbit}.
Its present velocity is summarized in \ta{tab:orbitdata}, with 
$U$ in the Galactic disk, positive to the Galactic center,
$V$ positive in the direction of Galactic rotation, and
$W$ pointing to the North Galactic pole (NGP).
We adopt the IAU convention \citep{kl1986} of $R_\odot = 8.5\,\mathrm{kpc}$ and
$V_{R_\odot} = 220\,\mathrm{km\,s^{-1}}$ for the Galactic rotational velocity at the Sun's position.
For the solar peculiar motion we adopt the values of \citet*{db1998} of
$U_\odot = 10.00\,\mathrm{km\,s^{-1}}$,
$V_\odot =  5.25\,\mathrm{km\,s^{-1}}$, and
$W_\odot =  7.17\,\mathrm{km\,s^{-1}}$.
The values of $U$, $V$, $W$, and of the Galacto-centric positions $X$, $Y$, $Z$, 
constitute the inputs to the code. Here, 
$X$ is positive in the direction of Galactic rotation, 
$Y$ positive from the Galactic center to the Sun, and 
$Z$ positive towards the NGP.

\begin{table}
\caption[]{Data related to the orbit of \lsv. $z$ is the present height above the Galactic plane, 
the given angle is measured between the present orbit direction and the Galactic latitude.
}
\label{tab:orbitdata}
\begin{tabular}{cccccccc} \\ 
\hline
\hline
$z$ & 
angle & 
$U$ & 
$\sigma(U)$ & 
$V$ & 
$\sigma(V)$ & 
$W$ & 
$\sigma(W)$ \\ 
\cline{3-8}\noalign{\smallskip}
pc & 
$\degr$ & 
\multicolumn{6}{c}{$\mathrm{km\,s^{-1}}$} \\ 
\hline
\noalign{\smallskip}
1 & 
2 &  
$-$14.2 &   
1.5 & 
219.0 &   
15.0 & 
12.5 &   
0.7 \\
\hline
\end{tabular}
\end{table}

\section{Is \pnsh\ a planetary nebula?}
\label{sect:truepn}

\lsv\ is definitely a post-AGB star with a mass of $M = 0.550\,\mathrm{M_\odot}$ \sK{subsect:MdL}.
During its AGB mass-loss phase, it has lost about 75\% of its initial mass into the ISM.
Due to an interaction with the ambient ISM, the previously ejected envelope matter began to
slow down about 45\,000 ago and \lsv\ is moving out of the geometric center. 
However, it is still surrounded by its ``own'' material and ionizes it. 
Narrow-band images of \pnsh\ exhibit a shell-like nebula.
Its expansion time can not be calculated reliably because the measured upper limit for the present
$v_\mathrm{exp}$ of 4\,km\,s$^{-1}$ \citep{r1985} has a relatively large error range. Moreover, the low
$v_\mathrm{exp}$ is the result of interaction with the ISM and, thus, the rate of change in that
velocity over time is difficult to extrapolate back in time. If we assume a constant
$v_\mathrm{exp} = 4\,\mathrm{km\,s^{-1}}$, we get an upper limit for expansion time of about
460\,000 years.
This expansion time is in agreement with evolutionary calculations of \citet{s1983} for the stellar
remnant. However, the real expansion time may be much smaller.

The classification of \pnsh\ as a PN is corroborated but not proved, however,
based on the latest parameters of \lsv\ and \pnsh\, e.g\@. the determination of
the ionized nebula mass could give evidence for \pnsh\ being a ``real'' PN.
We can calculate that, at a distance of 128.9\,pc \citep{hea2007}, an apparent diameter of 100' gives a linear
radius of $r_\mathrm{sh} = 1.87\,\mathrm{pc}$. If we assume that an old nebular shell has a thickness
of $\approx 0.02 r_\mathrm{sh}$, assume a density of five particles / cm$^3$ \citep[cf.][]{tmn1995}
and a solar composition ($\mu = 1.26$), then we arrive at $m_\mathrm{sh} \approx 0.2\,\mathrm{M_\odot}$.
Although this may be a coarse approximation, the $m_\mathrm{sh}$ appears in agreement with an
``average'' PN mass.

\section{Conclusions}
\label{sect:conclusions}

The successful reproduction of the high-resolution FUSE and STIS spectra of \lsv\ by our
synthetic spectra calculated from NLTE model atmospheres shows that 
---  when done with sufficient care --- theory works.

\begin{figure}[ht]
  \resizebox{\hsize}{!}{\includegraphics{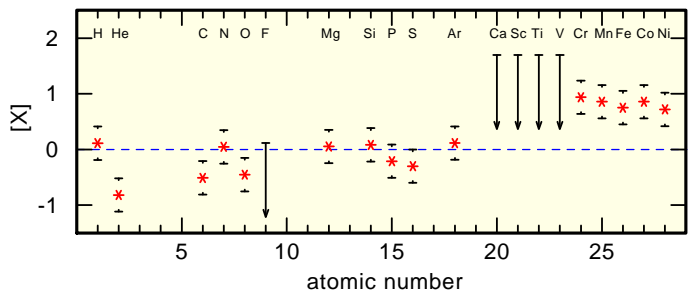}}
  \caption[]{Photospheric abundances of \lsv\ determined from detailed line profile fits. 
             For F, Ca, Sc, Ti, and V upper limits can be found only.
            }
  \label{fig:abundances}
\end{figure}

From the 
\Ion{N}{4} -- \Ion{N}{5}, 
\Ion{O}{4} -- \Ion{O}{6}, 
\Ion{Si}{4} -- \Ion{Si}{5}, and
\Ion{Fe}{5} -- \Ion{Fe}{7} 
ionization equilibria, we were able to determine
\Teffw{95\pm 2} with -- for these objects -- unprecedented precision. Since this is a prerequisite
for reliable abundance determinations (Tab\@. \ref{tab:abundances}), the error limits for the determined abundances  
became also correspondingly small (Fig\@. \ref{fig:abundances}). 

The previously determined surface gravity of \loggw{6.9\pm 0.2} \citep{tea2005} is confirmed. 
Thus, the spectroscopic distance of 224\,pc \sK{subsect:MdL} is too large compared with the
parallax distance of 129\,pc \citep{hea2007}.
It is disconcerting that we have such a large discrepancy for this star that is so well observed.  
\Teff\ is well determined by different ionization equilibria and, thus, only a higher
\logg\ of about 7.4 \sK{subsubsect:g} could reconcile the discrepancy since the parallax distance
requires a smaller stellar radius. 

\citet{n2001} investigated distance scales of old PN and found that trigonometric distances ($d_\Pi$) are
always smaller than spectroscopic distances determined by means of NLTE models ($d_\mathrm{NLTE}$).
He found a distance ratio of
$r_\mathrm{obs} = d_\mathrm{NLTE}/d_\Pi = 1.55 \pm 0.29$.
However, simulations showed that a significant fraction of this discrepancy could be explained
by biases caused by large relative errors of some parallax measurements \citep[cf\@. Fig\@. 2 in][]{n2001}.
Recently, \citet{hea2007} used improved error estimates for both distances and determined $r_\mathrm{obs} = 1.3$.
The parallax errors of several objects are significantly reduced and so are the biases as discussed in \citet{hea2007}.

In the case of \lsv\, we arrive at a ratio of $r_\mathrm{obs} = 1.74 \pm 0.35$ \citep[cf.][$r_\mathrm{obs} = 1.46$]{n2001}
which is in accordance with the apparently general problem.
However, one of the aims of this work was to solve this problem \citep[cf.][and the discussions therein]{n2001,hea2007}
-- but is remains at almost the same level although we use very elaborate NLTE model atmospheres. 
Since the {\it TMAP} NLTE model atmospheres \sK{subsect:model} used for this analysis are homogeneous, 
it would be highly desirable to calculate stratified models for \lsv\ in order to search for significant differences.

\begin{table}[ht]
\caption[]{Photospheric abundances of  \lsv\ normalized to 
$\log \sum_i\mu_i\varepsilon_i = 12.15$,
compared with solar abundances \citep{ags2005}.
} 
\label{tab:abundances}
\begin{tabular}{cr@{.}lr@{.}l}
\hline
\hline
\noalign{\smallskip}
element & \multicolumn{2}{c}{$\log \varepsilon$} & \multicolumn{2}{c}{[x]} \\
\hline
 H   & 12 & 035 &    0 & 112 \\
 He  & 10 & 634 & $-$0 & 818 \\
 C   &  8 & 880 & $-$0 & 510 \\
 N   &  8 & 892 &    0 & 046 \\
 O   &  9 & 332 & $-$0 & 453 \\
 F   &  5 & 875 & $<$0 & 116 \\
 Mg  &  8 & 889 &    0 & 053 \\
 Si  &  8 & 960 &    0 & 082 \\
 P   &  6 & 560 & $-$0 & 211 \\
 S   &  8 & 265 & $-$0 & 301 \\
 Ar  &  7 & 781 &    0 & 079 \\
 Ca  &  8 & 583 & $<$1 & 750 \\
 Sc  &  5 & 442 & $<$1 & 819 \\
 Ti  &  7 & 320 & $<$1 & 819 \\
 V   &  6 & 327 & $<$1 & 699 \\
 Cr  &  8 & 214 &    0 & 938 \\
 Mn  &  7 & 908 &    0 & 858 \\
 Fe  &  9 & 867 &    0 & 750 \\
 Co  &  7 & 468 &    0 & 858 \\
 Ni  &  8 & 638 &    0 & 719 \\  
\hline
\end{tabular}
\end{table}

The derived abundance pattern (Fig.~\ref{fig:abundances}) gives evidence for an interplay of
gravitational settling (e.g\@. the He and C abundances are strongly decreased by a factor of $\approx 0.15$) 
and radiative levitation (iron-group elements show an abundance up to $\approx 10\times$ solar).

A comparison to abundance predictions \sA{fig:diffusion} from diffusion calculations of \citet*{cfw1995} and \citet*{cea1995} 
shows a good agreement for C, N, O, Mg, S, and Fe with the DA values 
while we find that Si has a higher abundance and Ar and Ca have a lower abundance in our model. 

Diffusion calculations for He in DAO white dwarfs have been presented by \citet{vea1988}. These predict
an upper limit of $\log \mathrm{He/H} \approx -2.5$ for a $0.55\,\mathrm{M_\odot}$ and \Teffw{95} star.
We arrive at $\log \mathrm{He/H} = -2.0$ in our final model.

New diffusion calculations which include all elements from hydrogen to nickel could further improve our models,
e.g\@. provide abundance predictions for those species which are not identified in the spectrum of \lsv.

\begin{figure}[ht]
  \resizebox{\hsize}{!}{\includegraphics{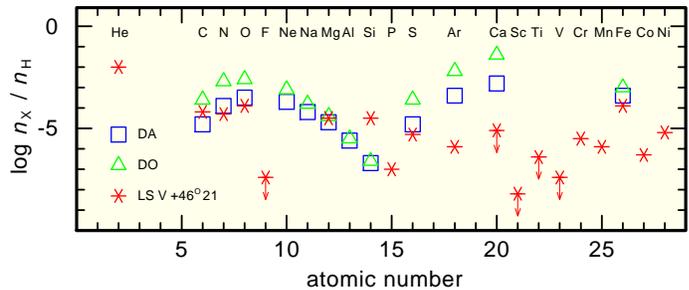}}
  \caption[]{Comparison of the elemental  number ratios found in our spectral analysis compared to predictions of
             diffusion calculations for DA and DO models \citep{cea1995, cfw1995} with \Teffw{95} and \loggw{7}. 
            }
  \label{fig:diffusion}
\end{figure}

\citet*{gea2005} have determined C, N, O, Si, and Fe abundances using FUSE observations of 16 DAO white dwarfs.
They compare their results with values given by \citet{cfw1995} and find deviations of
$+0.25$, $-0.50$, $-0.10$, $+3.0$, and $+0.1$\,dex (interpolated by us in their figures at \Teffw{95}), respectively. 
This is within 0.5\,dex in agreement with our result.

In our metal-line blanketed NLTE model atmospheres which include the opacity of 20 species
from H to Ni \sK{sect:modeling},
the BLP \sK{sect:introduction} vanishes \sA{fig:blp2006} if the synthetic spectra are compared
to the available medium-resolution optical spectra at $S/N \approx 30$. High-resolution and
high-$S/N$ spectra are highly desirable to check for remaining discrepancies.

The reddening of $E_\mathrm{B-V} = 0.065^{+0.010}_{-0.015}$ towards \lsv\ is much higher than expected from the
Galactic reddening law (Sect.~\ref{subsect:ISMbasic}) possibly because of additional reddening due to dust in the
nebula \pnsh.

\section{Future Work: Atomic and line-broadening data}
\label{sect:future}

With the analysis of the high-resolution, high-S/N UV spectrum of \lsv\ it is demonstrated that
state-of-the-art NLTE spectral analysis has presently arrived at a high level of sophistication.
The main limitation now encountered is the lack of reliable atomic and line-broadening data.

Going to higher resolution and S/N in the observations reveals uncertainties 
in atomic data even for the most abundant species.
Thus, it is a challenge for atomic physics to provide properly measured atomic data 
(for a wide range of ionization stages).

Discrepancies between our synthetic line profiles of the resonance doublets of \Ion{N}{5},
\Ion{O}{6}, and \Ion{S}{6} and the observation may have their reason in insufficient 
line-broadening tables. Even for strategic lines like, e.g., \Ionw{O}{5}{1371}
which is often used to determine \Teff,
no appropriate tables are available. 
Theoretical physics should provide data which covers the astrophysical relevant temperature and density space.

Better atomic and line-broadening data
will then strongly improve future spectral analyses and thus, make determinations of
photospheric properties more reliable.

\begin{acknowledgements}
We like to thank Hugh Harris and Ralf Napiwotzki for comments and discussion.
T.R\@. is supported by the BMBF/DESY under grant 05\,AC6VTB.
J.W.K\@. is supported by the FUSE project, funded by NASA contract NAS532985.
This research has made use of the SIMBAD Astronomical Database, operated at CDS, Strasbourg, France.
This work has been done using the profile fitting procedure {\sc Owens} developed by M\@. Lemoine and the FUSE French Team.
\end{acknowledgements}

\bibliographystyle{aa}
\bibliography{7166.bbl}

\end{document}